\documentclass[prr,twocolumn,amsmath,amssymb]{revtex4-2}
\usepackage{graphicx}
\usepackage{amssymb}
\usepackage{amsmath}
\usepackage[braket, qm]{qcircuit}
\usepackage{bbold}
\usepackage{appendix}

\newcommand{\nmax}{n}
\usepackage{bbm}
\usepackage{lineno}
\usepackage{slashbox}
\usepackage{color}
\usepackage{hyperref}
\begin{document}
\title{%Experimental l
Lower bounds on entanglement entropy without twin copy}

\author{Yannick Meurice}
\affiliation{Department of Physics and Astronomy, The University of Iowa, Iowa City, IA 52242, USA }
%\date{February 2022}

 \date{\today}

\begin{abstract}
We discuss the possibility of estimating experimentally the von Neumann entanglement entropy $S_{A}^{vN}$ of a symmetric bi-partite quantum system $AB$ by using the basic measurement counts (bitstrings)
for a {\it single} copy of a prepared state. Using exact diagonalization and analog simulations performed with the publicly available QuEra facilities for chains and ladders of Rydberg atoms, we calculate the Shannon entropy $S_{AB}^X$ associated with the bitstrings of adiabatically prepared ground  states and the reduced entropies  $S_A^X$ and  $S_B^X$  obtained from the marginal probabilities in $A$ and $B$. We then calculate the classical mutual information $I^X_{AB}=S_A^X+S_B^X-S_{AB}^X$, which is a lower bound on  $S_{A}^{vN}$. 
We show that for a broad range of lattice spacing and detuning, $I^X_{AB}$ is typically 20 percent below $S_{A}^{vN}$ in regions where $S_{A}^{vN}$ is large and a less close bound in regions where $S_{A}^{vN}$ is low. We argue that this use of the easily available bitstrings provides a robust and efficient way to explore empirically the phase diagram of qubit-based quantum simulators and identify critical regions.
%Similar results are found for the second order R\'enyi entanglement entropy.

 \end{abstract}

\maketitle

\def\beq{\begin{equation}}
\def\enq{\end{equation}}
\def\nq{n_q}
\def\nmax{n_{\mathrm{max}}}
\def\phix{\hat{\phi} _{\bf x}}
\def\nq{n_q}
\def\ah{\hat{a}}
\def\ahd{\hat{a}^\dagger}
%\ket{\nmax -1}\bra{\nmax -1}
\def\pn{P_{\nmax-1}}
\def\har{\hat{H}^{\mathrm{har.}}}
\def\hanh{\hat{H}^{\mathrm{anh.}}}
\def\hpf{\hat{\phi}^4}
\def\np{\mathcal{N}_p}

\vskip2pt\noindent
{\it Introduction}. Entanglement entropy is a  crucial theoretical concept in  quantum  physics with applications for many-body physics \cite{amico2008entanglement,Eisert:2008ur,Abanin_2019,Cirac:2020obd}, gauge theory \cite{Ghosh:2015iwa,VanAcoleyen:2015ccp,Banuls:2017ena,Knaute:2024wfh}, high-energy collisions \cite{Kharzeev:2017qzs, PhysRevD.98.054007,Zhang:2021hra}, nuclear physics \cite{Beane:2018oxh,Robin:2020aeh} and studies of conformal field theory \cite{PhysRevLett.90.227902,PhysRevLett.92.096402,Calabrese:2005zw,Ryu:2006bv}. It provides important information regarding quantum phases transitions.  
However, the experimental measurement of entanglement entropy is notoriously difficult due to its 
nonlocal nature. This led to the idea of coupling several
copies of the original system \cite{PhysRevLett.109.020504}. A standard proposal is to measure the second order  R\'enyi entropy $S_2$ by performing a swap operation between twin copies of a system \cite{PhysRevLett.109.020505}. Cold atom experimentalists \cite{Islam:2015mom,Kaufman:2016mif} were able to prepare and interfere
twin copies of a state in small optical lattices to measure $S_2$. If practically feasible for larger systems, this would allow the experimental measurement of the central charge which is an important characterization of  conformal systems \cite{PhysRevLett.90.227902,PhysRevLett.92.096402,Calabrese:2005zw,Unmuth-Yockey:2016znu}.

Preparing and interfering twin copies of a quantum system are not easy experimental tasks. On the other hand,  a generic qubit-based quantum computing device provides %experimental 
counts for bitstrings often given as python dictionaries. For instance, for  1000 shots  with a two-qubit universal 
quantum computer one gets counts for the four possible states in the form \verb|{'01': 269, '00': 251, '10': 247, '11': 233}|.  As another example, for a 10 Rydberg atom analog simulator where $g$ and $r$ represent, for each atom,  the ground and Rydberg states respectively the counts are given as \verb|{'gggrgrgrgr': 8, 'rgrgrggrgr': 160,  ...}|. By dividing these counts by the total number of shots, we can estimate the probabilities $p_{\{n\}}$ for the bitstring states $\ket{\{n\}}$. We can then introduce an $AB$  bipartition and calculate the marginal probabilities in $A$ or $B$. For instance in the 10 atom example, we can estimate the probability to observe \verb|grgrgr| on the right side of a chain irrespectively of the observations made in $B$. This reduction process is reminiscent of the tracing over a subsystem used to calculate the von Neumann quantum entanglement $S^{vN}_A$ which is equal to $S^{vN}_B$ for a pure state.

In this Letter, we discuss the applications of concepts introduced by Shannon \cite{shannon49} in the context of the transmission and interpretation of classical bits, to the bitstring measurements of bipartite quantum systems. 
Using the Shannon entropy associated with these bitstrings and reduced versions  obtained from the marginal probabilities,  it is easy to calculate the classical mutual information $I^X_{AB}$, which is a lower bound on  $S_{A}^{vN}$ \cite{nandc,jpch10}. 
We show that for a broad range of tunable parameters, $I^X_{AB}$ is typically 20 percent below $S_{A}^{vN}$ in regions where $S_{A}^{vN}$ is large and a less close bound in regions where $S_{A}^{vN}$ is low.
In other words, $I^X_{AB}$
provides unexpectedly good characterizations of the variations of $S^{vN}_A$ and allows an efficient exploration
of phase diagrams. 
The bitstrings  are available on many analog and universal quantum computing platforms.
In the following, we focus on configurable arrays of Rydberg atoms.  We used exact diagonalization and conducted 
%experimental 
analog simulations using the QuEra device Aquila \cite{wurtz2023aquila} which is publicly available and can be operated with limited resources.

The presentation is organized as follows. We introduce the bitstring Shannon entropy and mutual information for a bipartite quantum system and then perform calculations for a chain of 10 Rydberg atoms with a fixed detuning to Rabi frequency ratio $\Delta/\Omega$  and a variable lattice spacing. We then extend the numerical calculations for a broad range of $\Delta/\Omega$ and more general situations. 

{\it Bi-partite setup, bitstring mutual information and quantum bounds}. In the following we focus on the vacuum density matrix $\rho_{AB}=\ket{vac.}\bra{vac.}$ of a bipartite $AB$ quantum system. We define 
the reduced  density matrix $\rho_A={\rm Tr}_B\rho _{AB}$  and the von Neumann entanglement entropy 
\beq 
\label{eq:vn}
S_{A}^{vN}\equiv-{\rm Tr}(\rho_A \ln(\rho_A)), \enq which is independent of the bases used in $A$ and $B$. 
We now express $\rho_{AB}$ in the computational basis which has a bipartite factorization $\ket{\{n\}_{AB}}=\ket{\{n\}_A}\ket{\{n\}_B}$. 
In this basis, the diagonal elements are the probabilities for the bitstrings.
% and by discarding the non-diagonal element, 
%We define 
 The Shannon entropy for the experimental probabilities is defined as:
\beq
S_{AB}^{ X}\equiv -\sum_{\{n\}} p_{\{n\}}\ln(p_{\{n\}}).
\enq
The superscript $X$ is short for eXperimental as opposed to quantum von Neuman (or R\'enyi).

Similarly, we can consider the diagonal part of $\rho_A$. Since the partial tracing over $B$ is a diagonal operation, the reduced diagonal elements 
\beq
p_{\{n\}_A}=\sum_{\{n\}_B}p_{\{n\}_A \{n\}_B},
\enq
are the marginal probabilities introduced in the 18th century and we define the reduced experimental entropy 
\beq
S_{A}^{X}\equiv -\sum_{\{n\}_A} p_{\{n\}_A}\ln(p_{\{n\}_A}).
\enq
This quantity is basis dependent and its relation to the basis independent $S_{A}^{vN}$ can in general be expressed as an upper bound using convexity arguments \cite{jpch10,wittenmini} :
\beq
S^{vN}_A\leq S_A^X
\enq
The process can be repeated for $\rho_B$ leading to $S_B^X$. In general, $S_B^X\neq S_A^X$ as it can be easily seen by taking $A$ and $B$ with very different sizes. This is contrast to the quantum symmetry $S_B^{vN} = S_A^{vN}$ for a pure state.% density matrix.

Shannon \cite{shannon49} introduced a manisfestly $A-B$ symmetric, non-negative, quantity usually called the mutual information $I_{AB}$ which is a measure of the information shared by $A$ and $B$. In our bitstring context, it  reads:
\beq
I^X_{AB}=S_A^X+S_B^X-S_{AB}^X. 
\enq
It is clear that in the limit where $p_{\{n\}}=p_{\{n\}_A} p_{\{n\}_B}$, the measurements in $A$ and $B$ are completly unrelated and $I_{AB}^X=0$.
At the quantum level, if the pure state is a product of two states in $A$ and $B$, then $S_B^{vN} = S_A^{vN}=0$. This suggests that there could exist a relation between $I^X_{AB}$ and $S_A^{vN}$. We first empirically 
observed that $I^X_{AB}$ was keeping track of $S_A^{vN}$ from below and 
it was pointed out to us \cite{mehdi} that a general inequality can be obtained from the Holevo bound discussed in \cite{nandc,jpch10}:
\beq
I^X_{AB}\leq S^{vN}_A=S^{vN}_B
\enq

In summary, we have provided upper and lower bounds on the von Neumann entanglement that can be easily obtained from bitstring data. 
The main question addressed hereafter is to decide if some of these bounds can be tight enough to provide reliable information about quantum entanglement. 

In the rest of this Letter, we  assume that $A$ and $B$ have the same size and can be interchanged by a reflection symmetry. This implies that $S_A^X=S_B^X$ and $I_{AB}^X=2S_A^X-S_{AB}^X$. 
Applications in assymmetric cases work well and  are discussed elsewhere \cite{avi,zaneprogress}.

{\it Rydberg atoms setup}. Recently, optical tweezer have been used to create arrays of Rydberg atoms with adjustable geometries \cite{Bernien2017Dynamics, Keesling2019Kibble, Labuhn2016RydIsing,Leseleuc2019topo,Ebadi2021_256,Pascal2021AF}. 
Their  Hamiltonian reads

\beq
\label{eq:genryd}
H = \frac{\Omega}{2}\sum_i(\ket{g_i}\bra{r_i} + \ket{r_i}\bra{g_i})-\Delta\sum_i  n_i +\sum_{i<j}V_{ij}n_in_j,
\enq
with van der Waals interactions $V_{ij}=\Omega R_b^6/r_{ij}^6$,
for a distance $r_{ij}$ between the atoms labelled as $i$ and $j$.  By definition, $r_{ij}=R_b$, the Rydberg blockade radius,  when $V_{ij}=\Omega$ the Rabi frequency. We define the Rydberg occupation  $n_i\ket{r_i}=\ket{r_i}$ while $n_i\ket{g_i}=0$. In the following we use $\Omega=5\pi$ MHz, which implies $R_b=8.375 \mu$m, and a detuning $\Delta=17.5\pi$ MHz as in \cite{floating}. We will consider chains and ladders of atoms with varying lattice spacings.  The adiabatic preparation is discussed in Apprendix \ref{sec:exp} and the errors on the bitstring measurements in Appendix \ref{sec:errors}. 

{\it Numerical calculations}. As a first step to understand the connections among the bitstring entropies and  $S_A^{vN}$, we consider the ground state of a chain of 10 Rydberg atoms separated by a varied lattice spacing $a_x$ expressed in units of $R_b$. $A$ and $B$ are the left and right sides with 5 atoms each. We used exact diagonalization to calculate $S_{AB}^X$, $S_A^X$, $I_{AB}^X$, and $S^{vN}_A$ for the vacuum. The results are shown in Fig. \ref{fig:3echain}. 
\begin{figure}[h]
\includegraphics[width=8.6cm]{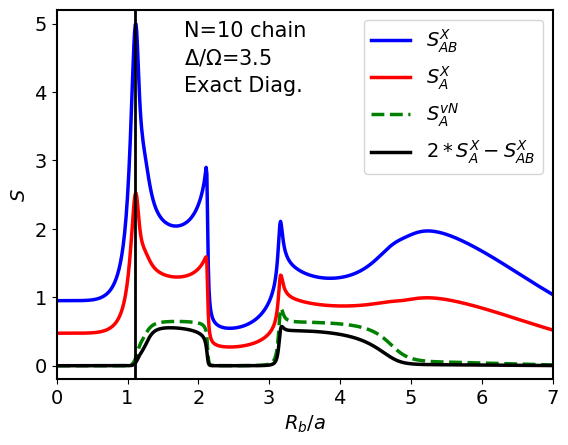} 
\caption{\label{fig:3echain} Entropies for a chain of 10 atoms with $\Delta/\Omega=3.5$ as a function of $R_b/a_x$ : $S_{AB}^X$, $S_A^X=S_B^X$, $I_{AB}^X=2S_A^X-S_{AB}^X$, and $S^{vN}
_A$; the vertical line is at $R_b/a_x=1.11$; }
\end{figure}

We observe that $S_{AB}^X$ has approximately the same shape as $S_A^X$ with a constant of proportionality which is  approximately 2. In addition 
$S_{A}^{vN}\simeq 1.25(2S_A^X-S_{AB}^X)$. As shown in the SM, the proportionality factor is only approximate, for instance, $S_{A}^{vN}$ is slightly below (above)  this linear combination for smaller (larger) $R_b/a_x$ respectively.  Notice that as $R_b/a_x$ is increased from low values, the onset of $S_{A}^{vN}$ is marked by a sharp increase $S_{AB}^X$ and $S_A^X$. However, at the peak of  $S_{AB}^X$ and $S_A^X$, for $R_b/a_x\simeq 1.11$, cancellations occur and $S_{A}^{vN}$ develops only after $S_{AB}^X$ and $S_A^X$ drop significantly. As expected, $S_A^X\geq S_{A}^{vN}$ but typically $S_A^X>>S_{A}^{vN}$ so this upper bound is not very useful.
Other features of Fig. \ref{fig:3echain} can be understood from elementary considerations that are provided in the SM  \cite{SM}. In summary, we found that whenever 
$S_{A}^{vN}$ is not too small, $I_{AB}^X\simeq 0.8 \times S_{A}^{vN}$ which is 20 percent below the target value.  When
$S_{A}^{vN}$ is small, $I_{AB}^X$ provide a bound that is not so close.
Consequently, the mutual information provides a good estimation of the regions of the phase diagram where $S_{A}^{vN}$ gets large.

\begin{figure}[h]
\includegraphics[width=8.6cm]{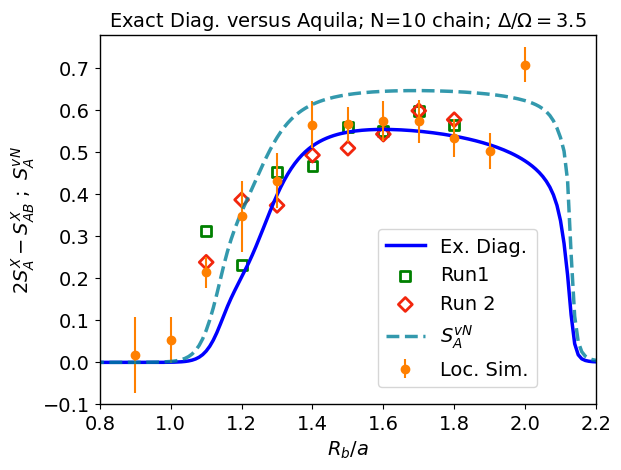} 
\caption{\label{fig:quera}  $2S_A^X-S_{AB}^X$ vs. $R_b/a_x$ with three methods: 1) exact diagonalization (continuous curve); 2) local simulator with no ramping down of $\Omega$ at the end (filled circles), the errors bars are calculated using 10 independent runs of 1000 shots; 3) 2 runs with 1000 shots with Aquila (empty symbols);  $S_{A}^{vN}$  (dashed line) is also given for reference. } 
\end{figure}

{\it Analog simulations}. We have repeated parts of the above numerical calculations using the empirical probabilities provided by the Aquila device and the associated Amazon Braket local simulator.  As explained in Appendix \ref{sec:exp}, this imposes size constraints not present in exact diagonalization.   Calculations for $1.0\lesssim R_b/a_x \lesssim 2$ can be performed with the local simulator and, for a slightly shorter range of $R_b/a_x$, with Aquila. In both cases, we ramped up the vacuum adiabatically starting with all atoms in the atomic ground state, increasing $\Omega$ and then $\Delta$, exactly as in \cite{floating} and illustrated in Appendix \ref{sec:exp}. The data is publicly available \cite{zenodo}. The results are shown in 
Fig. \ref{fig:quera} where we see good estimates of $2S_A^X-S_{AB}^X$  with Aquila. 
The local simulator becomes less accurate when $R_b/a_x \simeq 2$ and the ramping down of $\Omega$ at the end involves subtleties discussed in the SM \cite{SM}. Fig. \ref{fig:quera} only shows results with no ramping down of $\Omega$ at the end for the local simulator. 

In Appendix \ref{sec:errors}, we explain that the Aquila values for both $S_{AB}^X$ and $2S_A^X$ are significantly larger than the accurate numerical values, however these errors appear to cancel in $I_{AB}^X$. In the SM  \cite{SM}, we show that ``filtering" the low probabilities tend to increase $I_{AB}^X$ and get it closer to $S_A^{vN}$. This is discussed in more detail in Ref. \cite{avi}. 

{\it How tight is the lower bound?}  So far our model calculation shows that the mutual information $I^X_{AB}$ follows rather closely and from below the von Neuman entanglement entropy $S_{A}^{vN}$. This is not always the case. An example \footnote{This was pointed out to us by Alex Lukin.} of state $\ket{\psi}$ with  $I^X_{AB}=0$ and $S_A^{vN}=\ln 2$ is 
\beq
\label{eq:counter}
\ket{\psi}=\frac{1}{2}(\ket{gg}+\ket{gr}+\ket{rg}-\ket{rr}).
\enq
This state can be obtained by starting with a product state of two $(\ket{g}+\ket{r})/\sqrt{2}$  and then applying the entangling unitary transformation $\exp(i\pi n_1 n_2)$.  
All the $p_{\{n\}}$ for both states are 1/4 and the mutual information is zero, but $S_A^{vN}$ changes from zero to $\ln2$ when the sign of the coefficient of $\ket{rr}$ changes. 
A discussion of the exactly solvable two qubit case is given in the SM \cite{SM}.

\begin{figure}[h]
\includegraphics[width=8.6cm]{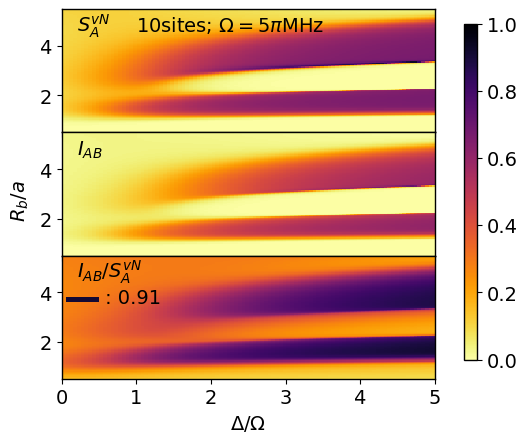} 
\caption{\label{fig:tight} $S^{vN}$, $I_{AB}^X$ and   $I_{AB}^X/S^{vN}$ for a 10 sites chain.}
\end{figure}

In order to get a better idea about the generic tightness of the bound, 
we have repeated our numerical calculations over a broad range of values of  $\Delta/ \Omega$. The results are shown in Fig. \ref{fig:tight}. 
We see that the shapes of the heatmaps of $S^{vN}$ and  $I_{AB}^X$ are very similar. The ratio $I_{AB}^X/S^{vN}$ reach maximal values of 0.91
% for the chain and 0.98 for the ladder
 in regions where 
$S^{vN}_A$ is large and small values in regions where $S^{vN}$ is close to 0. This confirms our previous observations that $I_{AB}^X$ delimitates reliably the regions of the phase diagram where  $S^{vN}_A$ takes its largest values.
We have repeated some of the numerical calculations discussed above with two-leg ladders with five rungs and varied the lattice spacings in both directions. We found very similar features that are 
reported in the  SM \cite{SM}.

 \noindent
 {\it Other bitstring proxies}. 
In Ref. \cite{floating}, we found a detailed correspondence between the contour plots in the $(\Delta/\Omega,R_b/a)$ plane of the peak height of the structure factors,
which can be reconstructed from the  $p_{\{n\}}$ only, and the contour plots of $S_A^{vN}$ which require the knowledge of the vacuum wave function. 
This motivated the idea of finding proxies of $S_A^{vN}$ that depends on the $p_{\{n\}}$ only. Besides $I_{AB}^X$,
one can also construct a
second order  R\'enyi version of the $S^X$ quantities defined above: 
\beq
S_{2AB}^{ X}\equiv -\ln \sum_{\{n\}} p_{\{n\}}^2, \ {\rm and}\ S_{2A}^{X}\equiv -\ln \sum_{\{n\}_A} p_{\{n\}_A}^2. 
\enq
As shown in the SM \cite{SM}, we found numerically that for the 10 atom chain:
\beq
\label{eq:mainrenyi}
S_{2A}^{R}\approx 2S_{2A}^X-S_{2AB}^X.
\enq
with $S_{2A}^{R}=-\ln({\rm Tr} \rho_A^2)$.
Another possibility is the configurational entanglement entropy. 
The connection between correlations and purity \cite{Lukin_2019} or the 
reconstruction of quantum states using a restricted Boltzmann machine based on single-basis measurements \cite{Torlai_2019} 
 could help develop a more general understanding of our observations. Ref. \cite{Lukin_2019} introduces another 
 proxy $C$ for $S_A^{vN}$ in the case of the the Bose-Hubbard model called the configurational entanglement  entropy. 
 Adapting their definition for the case where the particle number is not conserved and where there is no number entropy, and introducing a function of $C$ that reproduces $S_A^{vN}$ for a two-qubit Bell state we propose to use
 \beq
 LC \equiv \ln(1+
\sum_{\{n\}} |p_{\{n\}}-p_{\{n\}_A}p_{\{n\}_B}|).
\enq
This quantity provides good estimates of $S_A^{vN}$ but has a more intricated behavior which will be discussed separately.
%depends on the $p_{\{n\}}$ only and has similar limitations. Comparisons of the different approaches are under %study.

{\it Applications}.
Arrays of Rydberg atoms have been proposed for quantum simulation of lattice gauge theories \cite{Zhang:2018ufj,Surace:2019dtp,Notarnicola:2019wzb,Celi:2019lqy,Meurice:2021pvj, Fromholz:2022ymy,Gonzalez-Cuadra:2022hxt,
Heitritter:2022jik,Bauer:2022hpo,Halimeh:2023lid}. More specifically, ladder arrays have been considered as quantum simulators for scalar electrodynamics \cite{Meurice:2021pvj} and their phase diagram has been studied experimentally \cite{floating} and with effective Hamiltonian methods \cite{Zhang:2023agx}. These results show a very rich phase diagram and the need to study the regions of parameter spaces where the correlation lengths are much larger than the lattice spacing (continuum limits). This has possible relevance for the study of inhomogeneous phases and the Lifshitz regime of lattice quantum chromodynamics \cite{Pisarski:2019cvo, Kojo:2009ha}.  Entanglement entropy considerations play an important role in the tentative designs of hybrid algorithms for collider event generators \cite{Heitritter:2022jik}. 

{\it Generalizations.}
It is possible to extend our procedure for arbitrary $AB$ partitions where $A$ and $B$ have different sizes. This allows us to study the conjectured lower bound for multipartite entropies which are of great interest in the context of conformal field theory \cite{Headrick:2007km,Hayden:2011ag,Lin:2023pvl}. 
Ongoing calculations show that our proxy provide reasonably tight lower bounds for digitized $\phi^4$ models with qudits \cite{zaneprogress}, 
Ising models in curved space \cite{syrprogress} and larger Rydberg arrays than the ones considered here \cite{jamesprogress}. 

{\it Conclusions}. 
We have shown that the bitstring mutual information provides robust and reliable estimates for the order of magnitude of the von Neumann entanglement entropy. This works effeciently for a broad range of adjustable parameters available in arrays of Rydberg atoms. The method can be applied to any qubit based universal or analog quantum computing device.
This provides a quick way to map the phase diagram and identify regions of interest from the point of view of approaching  continuum limits with quantum simulators. 

\appendix
\section{Analog computation  methods}
\label{sec:exp}
\begin{figure}[h]
\includegraphics[width=8.6cm]{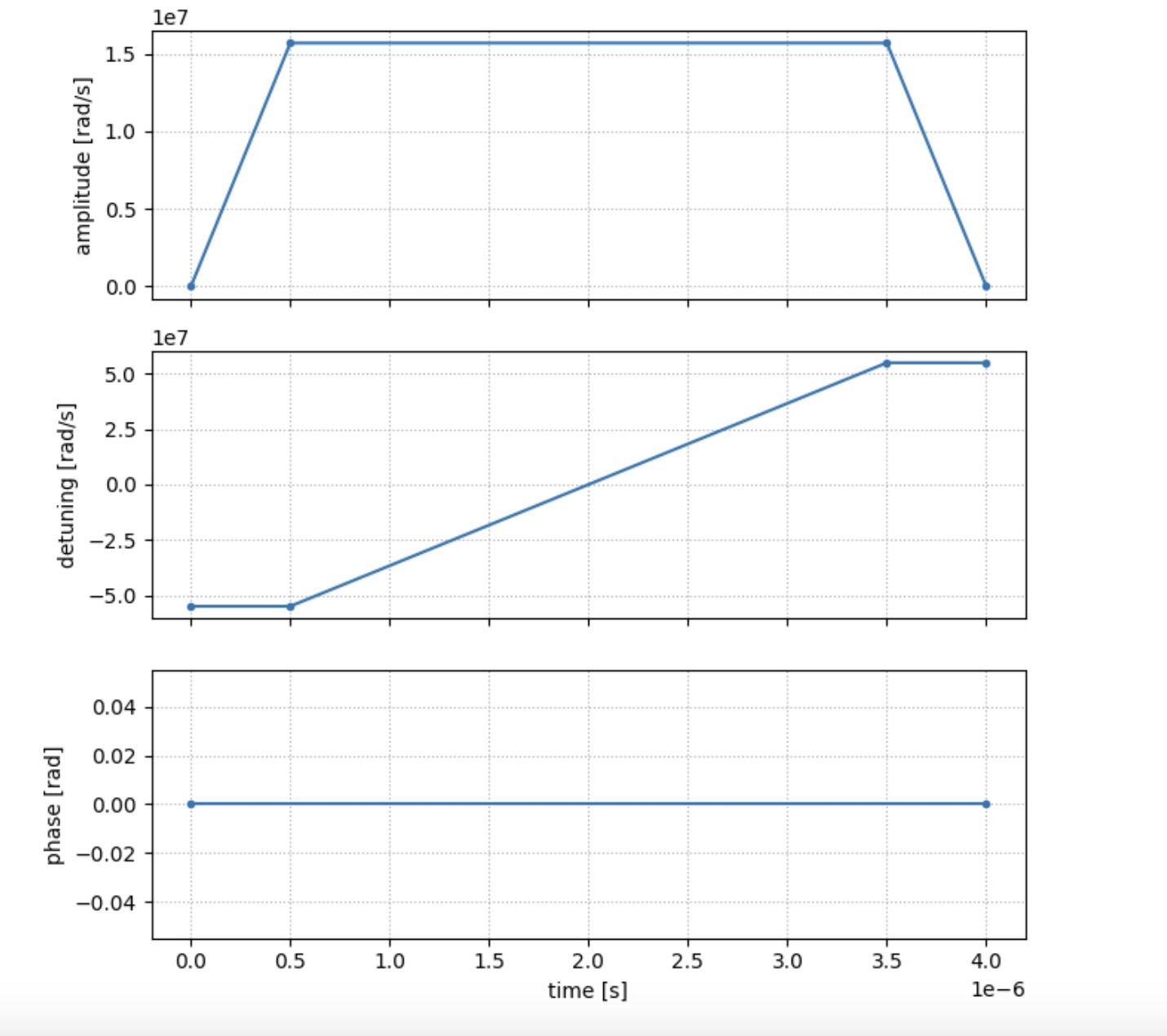} 
\caption{\label{fig:rampup} Standard ramping up or down of $\Omega$ (top), $\Delta$ (middle) and $\phi$ (bottom) as appearing in the Amazon Braket notebook tutorials.}
\end{figure}

Analog calculations with arrays of Rydberg atoms have been performed with Aquila using the Amazon Braket services. We used the Amazon Braket SDK for local simulations. The actual device imposes constraints not present in exact diagonalization. Analog calculations were performed with a single chain model with 10 sites, $\Omega=5\pi$ MHz, which implies $R_b=8.375 \mu$, and a detuning $\Delta=17.5\pi$ MHz. The linear size of the system needs to be less than $100\ \mu $m. This means that we need $R_b/a_x \gtrsim0.9$. The distance between  the atoms needs to be more than $4\ \mu$m which implies $R_b/a_x \lesssim 2.0$. We used the values $R_b/a_x=0.9, \ 1.0, \dots 2.0$. The ground state needs to be prepared adiabatically and actual measurements with Aquila are performed after turning off $\Omega$. The procedure suggested on the Amazon Braket tutorials  lasts 4 $\mu$s 
and involves ramping down $\Omega$ during the last 0.5 $\mu$s, as
illustrated in Fig. \ref{fig:rampup}. This protocole was used for our Aquila calculations. Different ramping down were studied with the local simulator and reported in the SM \cite{SM}.
We used 1000 shots 10 times for the local simulator and had  2 runs with 1000 shots for Aquila. The raw data files are available in \cite{zenodo}.

\section{Error cancellations}
\label{sec:errors}
\begin{figure}[h]
\includegraphics[width=4.2cm]{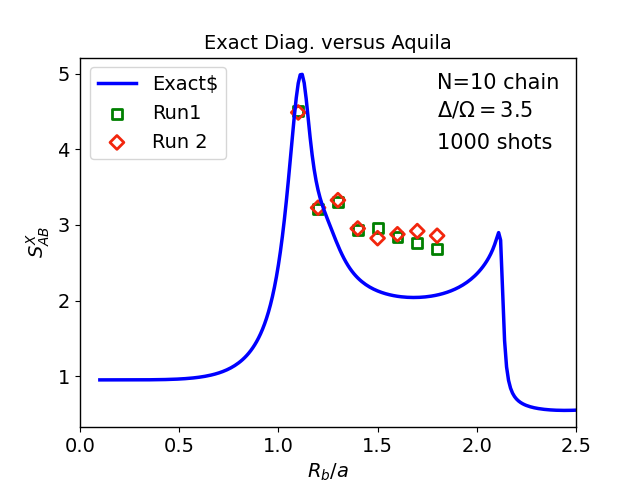} 
\includegraphics[width=4.2cm]{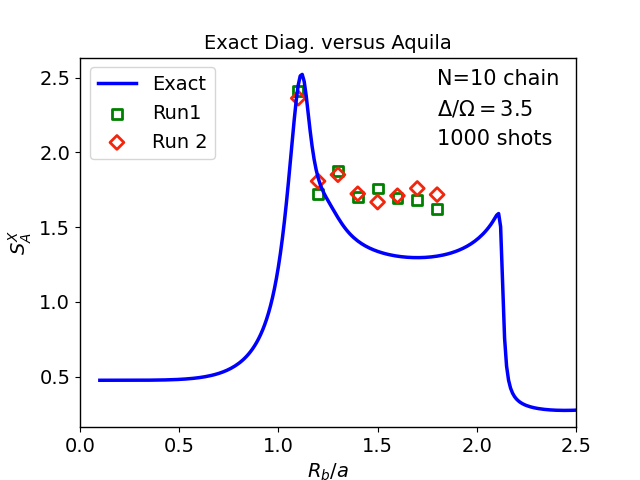} 
\caption{\label{fig:errors}Top: $S_{AB}^X$ (left), $S_A^X$ (right) vs. $R_b/a_x$ with exact diagonalization (continuous curves) and with Aquila with 2 runs of 1000 shots.} 
\end{figure}

An important aspect of the results presented in Fig. \ref{fig:quera} is that the errors (the difference between the analog estimates and the values calculated accurately with numerical methods) 
for $S_{AB}^{ X}$ and $2S_{A}^{ X}$ are significant and actually larger than $I_{AB}^X$.
This is illustrated in a specific way in Fig. \ref{fig:errors}. 
Near $R_b/a =1.5$, the difference between Aquila and exact diagonalization for both $2S_{A}^{ X}$ and $S_{AB}^{ X}$ are both positive and close to 1 while their difference $I_{AB}^X=2S_{A}^{ X} -S_{AB}^{ X}$ is of order 0.5 with an estimated error of order  0.1. More details can be found in the SM \cite{SM}. %In summary, the analog calculation of the bitstring mutual %information provides a good estimate of the order of magnitude of the quantum entanglement.

\section*{Acknowledgements} This research was supported in part by the Dept. of Energy
under Award Number DE-SC0019139. Special thanks to Alex Lukin who pointed 
out the counterexample of Eq. (\ref{eq:counter}) as well as the importance of understanding the proportionality factor.
We thank S. Cantu, M. Asaduzzaman, J. Corona, Cheng Chin, Sheng-Tao Wang, A. Bylinskii, Fangli Liu, J. Zeiher, R. Pisarski, P. Komar, B. Senseman, Z. Ozzello, Avi Kaufman and P. Preiss for comments and support. 
We thank J. Preskill and M. Soleimanifar for pointing out the connection between our empirical bound and the Holevo bound.
We thank 
the Amazon Web Services and S. Hassinger for facilitating remote
access to QuEra through the Amazon Braket while teaching quantum mechanics and our Department  of Physics and Astronomy for supporting the cost of the analog simulations presented here. The final version was completed while attending the workshop ``Co-design for Fundamental Physics in the Fault-Tolerant Era" at the 
InQubator for Quantum Simulation, University of Washington, Seattle, USA.
\begin{widetext}

\section*{Supplementary Material}
In this supplementary material we provide more information about: 

%I. Experimental methods

\hskip5pt  1: Symmetric bi-partite Hilbert spaces 

\hskip5pt  2: Detailed features of the entropies for chains with $\Delta/\Omega = 3.5$

\hskip5pt  3: Entropy errors 

\hskip5pt 4: Entropies for a 5 rung ladder and $\Delta/\Omega = 3.5$

\hskip5pt  5: Verification of the bound for two qubits 

\hskip5pt 6: Other bitstring proxies

 %\end{widetext}

%\section{Additional information}
\subsection{Symmetric bi-partite Hilbert space}
\label{app1}
%\section{Bi-partite setup}. 
In the following we consider $N_q$ two-state systems (qubits), for instance an array of  Rydberg atoms either in the ground $\ket{g}$ or Rydberg $\ket{r}$ state. For simplicity, we assume that $N_q$ is even and that the whole system ($AB$) can be divided in two subsystems $A$ and $B$ that can be mapped into each other by some reflection symmetry. A simple example is a linear chain of equally spaced Rydberg atoms with an even number of atoms. 

The computational basis consists of the $2^{N_q}$ elements 
\beq
\ket{\{n\}}\equiv\ket{n_0, n_1,\dots,n_{N_q-1}},
\enq
with $n_j=$ 0 or 1. Any element of this basis can be factored in a bi-partite way
with the two subsystems of identical size  A and B (each with $N_q/2$ qubits)
\beq
\ket{\{n\}_{AB}}=\ket{\{n\}_A}\ket{\{n\}_B},
\enq
with
\begin{eqnarray}
\ket{\{n\}_A}&\equiv&\ket{n_0, n_1,\dots,n_{N_q/2-1}} \ {\rm and } \\
\ket{\{n\}_B}&\equiv&\ket{n_{N_q/2},\dots,n_{N_q-1}}
\end{eqnarray}
Given an arbitrary prepared state $\ket{\psi}$, we can expand it in the computational basis
\beq
\ket{\psi}=\sum_{\{n\}} c_{\{n\}}\ket{\{n\}}, 
\enq
and the state $\ket{\{n\}}$ will be observed with a probability
\beq
p_{\{n\}}=|c_{\{n\}}|^2.
\enq
These probabilities can be estimated by measurements and define an ``experimental" entropy
\beq
S_{AB}^{ X}\equiv -\sum_{\{n\}} p_{\{n\}}\ln(p_{\{n\}}),
\enq
associated with the state $\ket{\psi}$. 
It is clear that this quantity depends on the computational basis and that it contains no information  about entanglement.

We now define a reduced probability in the subsystem $A$ by tracing over $B$:
\beq
p_{\{n\}_A}=\sum_{\{n\}_B}p_{\{n\}_A \{n\}_B},
\enq
and the corresponding reduced entropy
\beq
S_{A}^{X}\equiv -\sum_{\{n\}_A} p_{\{n\}_A}\ln(p_{\{n\}_A}).
\enq
Again, this quantity depend on the computational basis used and cannot be identified with the von Neuman entropy $S_A^{vN}$.

In the following we focus on the case where $\ket{\psi}=\ket{vac.}$, the vacuum of the Hamiltonian. Starting with 
\beq
\rho_{AB}=\ket{vac.}\bra{vac.}
\enq
and writing the $2^{N_q}$ dimensional vector $c_{\{n\}}$ corresponding to the vacuum as a $2^{N_q/2}\times 2^{N_q/2}$
matrix 
\beq
C_{\{n\}_A ,\{n\}_B} =c_{\{n\}}, 
\enq
we find that the reduced density matrix $\rho_A=Tr_B \rho_{AB}$
can be written as 
\beq
\rho_{A \{n\}_A ,\{n'\}_A} = (C C^{\dagger})_{\{n\}_A ,\{n'\}_A},
\enq
in the computational basis. The von Neuman entropy is
\beq
S_A^{vN}=-\sum_m \lambda_m \ln(\lambda _m), 
\enq
with $\lambda_m$ the eigenvalues of $\rho_A $ which are independent of the basis used in $A$. 

\subsection{Detailed features of the entropies for chains with  $\Delta/\Omega=3.5$}
\label{app2}

\begin{figure}[h]
\includegraphics[width=7.cm]{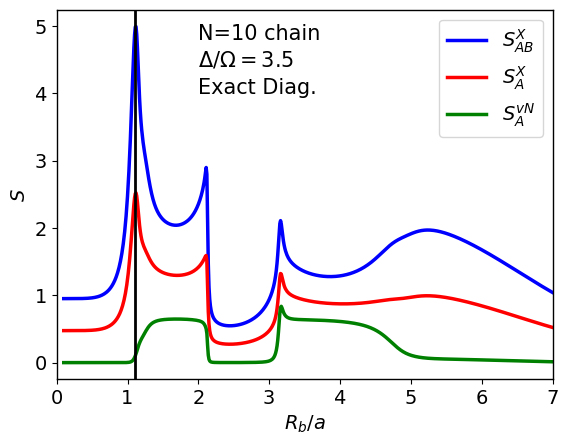} 
\includegraphics[width=7.2cm]{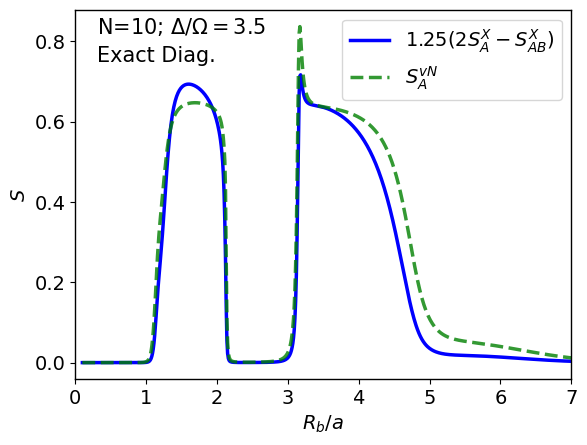} 
\caption{\label{fig:3echain} Entropies for a chain of 10 atoms with $\Delta/\Omega=3.5$ as a function of $R_b/a_x$. Top: $S_{AB}^X$, $S_A^X$ and $S^{vN}
_A$; the vertical line is at $R_b/a_x=1.11$; Bottom: $1.25(2S_A^X-S_{AB}^X)$ and $S^{vN}
_A$ .}
\end{figure}

Several features of Fig. \ref{fig:3echain} (here, Fig. 1 in the main text) can be understood from elementary considerations. For $R_b/a_x=0.5$, we have essentially 10 decoupled atoms. Each of the single-atom ground states have 
a probability $\cos(\theta/2)^2 =0.981$ (with $\theta=\arctan(\Omega/\Delta)$)  to be in the $\ket{r}$ state and the entropy per atom is 0.0951. Numerically we found $S_{AB}^X=0.9616\simeq 2S_A^X$ with 3 significant digits while $S_A^{vN}$ is negligible. Near $R_b/a_x$=1, $S_{AB}^X$ and $S_A^X$ increase rapidly to reach maxima near $R_b/a_x=1.1$
with $S_{AB}^X=4.874 \simeq 2 S_A^X =4.901$. The most probable state is $\ket{rrrrrrrrrr}$ with a probability of 0.036. 
For $R_b /a_x=1.5$, there are four states with probablities larger than 0.1: $\ket{rggrgrgrgr}$.  $\ket{rgrgrgrggr}$ (0.138) and $\ket{rgrggrgrgr}$ $\ket{rgrgrggrgr}$ (0.276). After tracing over the right part, there are three significant probabilities for $\ket{rggrg}$, 
 $\ket{rgrgg}$, and $\ket{rgrgr}$ and $2 S_A^X-S_{AB}^X=0.549$ while $S^{vN}_A=0.638$. For $R_b/a_x=2.5$, the state $\ket{rggrggrggr}$ has a probability 0.888 and both $2 S_A^X-S_{AB}^X$ and  $S^{vN}_A$ are small. 
 
 Fig. \ref{fig:chainbounds} shows that this inequality is verified for chains of size $N_s=2,\  4,\dots 10$. For $N_s=10$, the fraction of $S_{A}^{vN}$ not accounted by $2S_A^X-S_{AB}^X$ is compatible with the proportionality factor 1.25 used previously. 
\begin{figure}[h]
\includegraphics[width=7.cm]{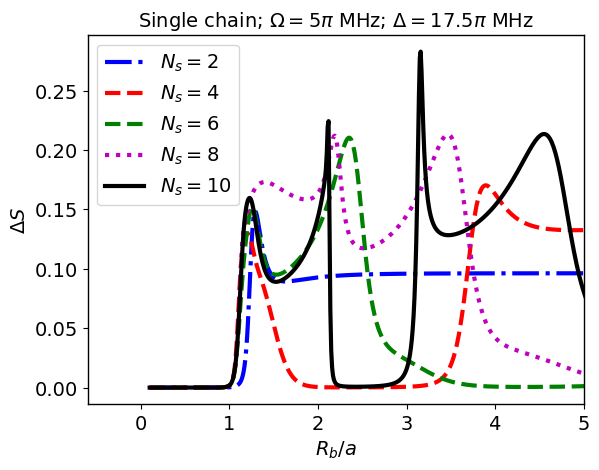} 
\caption{\label{fig:chainbounds} $S^{vN}-(2S_A^X-S_{AB}^X)$ as a function of $R_b/a_x$ for chains of size $N_s=2,\  4,\dots 10$.}
\end{figure}

\subsection{Entropy errors}
\label{app5}

In this supplemental information section, we compare the errors for the local simulator with no ramping down (LSNRD), the local simulator with the standard ramping down (LSST) as in Fig. 4 in the main text, and the actual device Aquila.

We started with the local simulator and found some dependence on the details of the ramping down of $\Omega$ at the end. The local simulator allows us to try time sequences not possible with the actual device. 
In order to check that the adiabatic preparation is working well, we first considered the situation with $\Omega(t)$ flat at the end (so $5\pi$ MHz after 4$\mu$ sec instead of ramping down to zero after 3.5 $\mu$s as in Fig. 4 in the main text).  We used 1000 shots and repeated the experiment ten times in order to estimate the statistical errors. The results are shown in Fig. \ref{fig:norampdown} and show good agreement with exact diagonalization. Except for $R_b/a_x=$ 1.1, 1.2 and 2.0, 
the values $1.25(2S_A^X-S_{AB}^X)$ obtained with the local simulator agree with less than one standard deviation with the exact diagonalization results. This gives us confidence in the adiabatic preparation of the ground state. 
\begin{figure}[h]
\includegraphics[width=7cm]{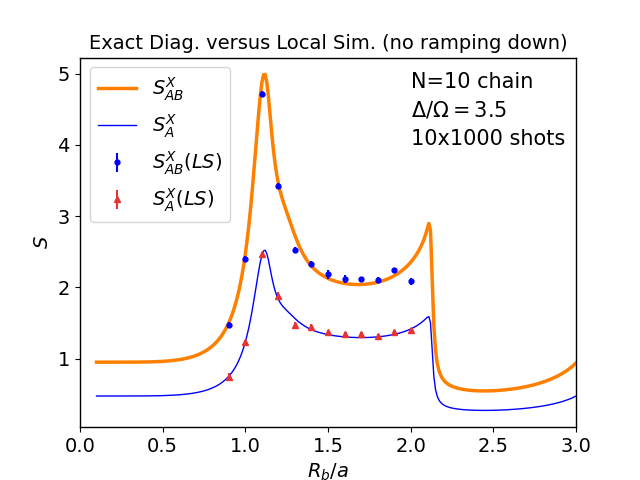} 
\includegraphics[width=7cm]{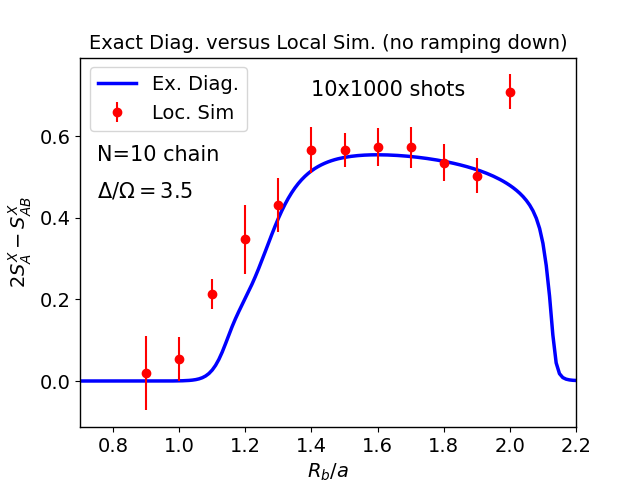} 
\caption{\label{fig:norampdown}Left:  $S_{AB}^X$ and $S_A^X$ vs. $R_b/a_x$.with exact diagonalization (continuous curves) and the local simulator with no ramping down of $\Omega$ at the end (symbols) using 1000 shots 10 times. The errors bars are present but barely visible on the graph. Right:  $1.25(2S_A^X-S_{AB}^X)$ with the same conventions. As the vertical scale is five times smaller, the errors bars are clearly visible.} 
\end{figure}

We repeated the local simulation with the realistic ramping down at the end shown in Fig. 4 in the main text. The results are shown in Fig. \ref{fig:withrampdown}. The agreement is not as good as in 
Fig. \ref{fig:norampdown}.
The size of the statistical errors bars are much smaller than the discrepancies with exact diagonalization for $S_{AB}^X$ and $S_A^X$. 
However the error somehow cancel in $2S_A^X-S_{AB}^X$ and we obtain reasonable estimates of $S_A^{vN}$. \begin{figure}[h]
\includegraphics[width=5cm]{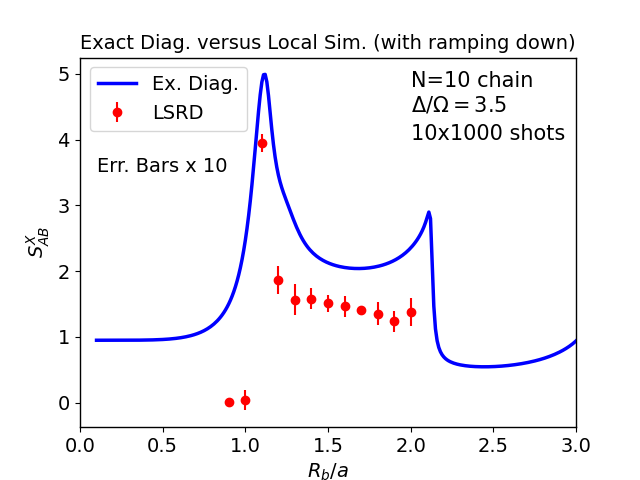} 
\includegraphics[width=5cm]{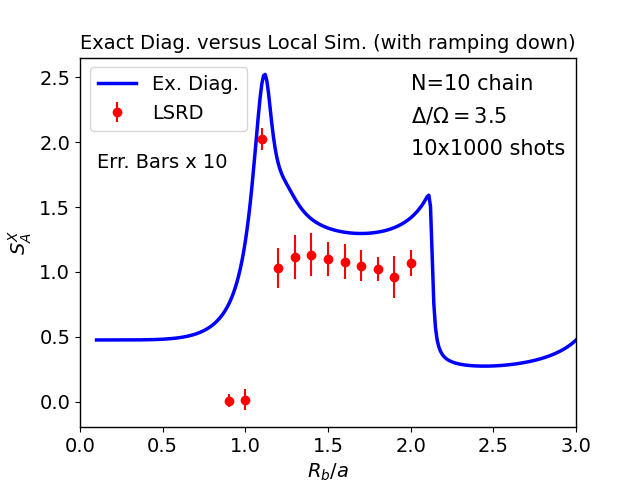} 
\includegraphics[width=5cm]{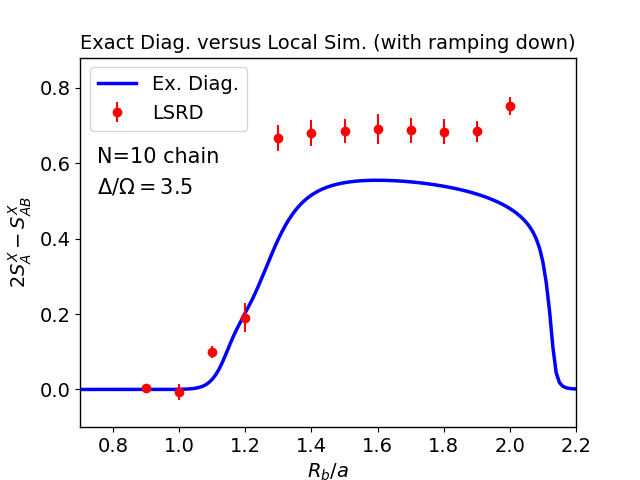} 
\includegraphics[width=5cm]{figs/entAquila1and2} 
\includegraphics[width=5cm]{figs/redentAquila.png} 
\includegraphics[width=5cm]{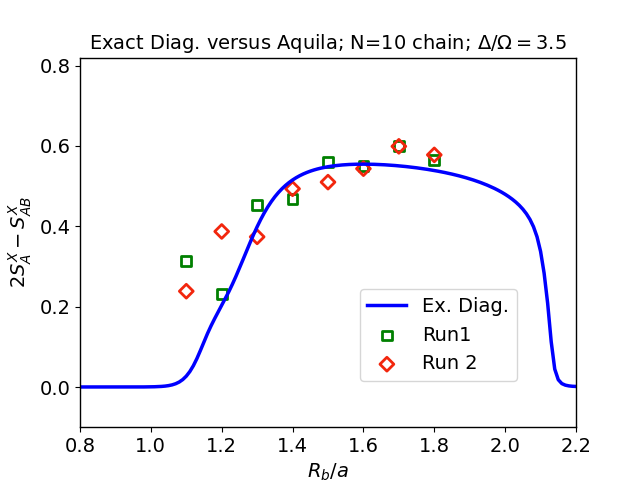} 

\caption{\label{fig:withrampdown}Top: $S_{AB}^X$ (left), $S_A^X$ (middle) and $2S_A^X-S_{AB}^X$ (right) vs. $R_b/a_x$ with exact diagonalization (continuous curves) and the local simulator with ramping down of $\Omega$ at the end  as in Fig. 4 in the main text (symbols) using 1000 shots 10 times. The errors bars are multiplied by 10. Bottom: same quantities with Aquila with 2 runs of 1000 shots.} 
\end{figure}
\begin{figure}[h]

\end{figure}

We repeated parts of these simulations with the actual device and we can now compare the estimations of $S_{AB}^X$ and $S_A^X$ obtained with: 1) exact diagonalization, 2) the local simulator with no ramping down (LSNRD) as in Fig. \ref{fig:norampdown}, 3) the local simulator with the standard ramping down (LSST) as in Fig. \ref{fig:withrampdown}, and 4) the actual device Aquila. We focus the discussion on $R_b/a_x=1.5$. 
In cases 2), 3) and 4) we used two independent runs with 1000 shots. By dividing the number of events by 1000, we obtain estimations of the probabilities that can be calculated exactly. 

The results involving more than 10 counts are shown in Table \ref{tab:states} and lead to the conclusions that:
\begin{itemize}
\item
LSNRD provides reasonably accurate estimations of the probabilities for states in the computational basis, for a broad range of values for the probabilities. This is not a surprise, it just shows that the local simulator works properly. 

\item
LSST removes states with low probabilities and significantly amplifies some medium probabilities such as 0.138 for rggrgrgrgr and rgrgrgrggr. This suggests that a more rapid ramping down maybe preferable for entropy estimation. 

\item
Aquila amplifies some of the low probability states and depletes higher probability states while keeping some ratios roughly in line with exact results. \end{itemize}

From the above discussion, it is clear that there are significant fluctuations for states that are expected to be seen 10 times or less in 1000 shots. LSST removes these states while Aquila amplifies them. 
This suggests to discard results with low counts and proceed with truncated data sets. 
In the following we considered truncated data sets where states with 10 or less measurements were discarded. By doing this, the number of shots is reduced which leads to different probability estimates. 
The results of this truncation for the entropies are shown in Table \ref{tab:ent}. From these results we conclude that

\begin{itemize}
\item
The truncation lowers the entropies calculated with exact diagonalization by about 20 percent but increases $1.25(2S_A^X-S_{AB}^X)$ by about 10 percent.
\item
LSNRD follows closely the exact results. Again this was expected. 
\item
LSST is insensitive to the truncation. The LSST values are close to exact truncated values. Again this was expected. 
\item
Aquila results are very sensitive to truncation which brings the results closer to exact ones, but maybe not as close as hoped. 

More generally, the truncation method seems to introduce significant uncertainties and we believe that error cancellations with the untruncated procedure may be more reliable. 

\end{itemize}

%\begin{widetext}

\begin{table}[h]
 \begin{tabular}{|c|c|c|c|c|c|c|c|}
\hline
\backslashbox{State}{Method}
& Exact&LSNRD1&LSNRD2&LSST1&LSST2&Aquila1&Aquila2\\
\hline
grgrgrgrgr &12&14&19&24&18&16&18\\
\hline
rggggrgrgr &16&21&18&$<10 \ (0)$&$<10 \ (1)$&15&14\\
\hline
rggrgrgrgr &138&142&152&201&222&136&149\\
\hline
rgrggggrgr &23&18&28&$<10 \ (0)$ &$<10 \ (0)$&19&24\\
\hline
rgrggrgrgg  &$<10 \ (6)$ &11&$<10 \ (7)$&$<10 \ (0)$ &$<10 \ (0)$&14& $<10$ (5)\\
\hline
rgrggrgrgr &276&232&239&274&264&174&175\\
\hline
rgrgrggggr &16&14&18&$<10 \ (0)$ &$<10 \ (0)$&19&12\\
\hline
rgrgrggrgr &276&248&244&278&266&190&180\\
\hline
rgrgrgrggr &138&179&168&207&200&151&179\\
\hline
rgrgrgrgrg &12&13&18&12&25&23&16\\
\hline
Total&907&892&904&996&995&804(*)&812(**)\\
\hline
\end{tabular}
\caption{\label{tab:states}States with at least 10 observations for 1000 shots. For Aquila, 
(*) five addtional states are not in the table:
ggrgrggrgr (12),  rgggrgrggr (12), rggrgrgggr (12),  rgrggrgggr (14),  rgrgrgrgrr (11)
for run 1, and 
(**) and three additional states are not in the table: rgggrggrgr (12), rggrgggrgr (11), rgrgrggrgg (12)  for run 2.}
\end{table}

\hfill 
\break

\begin{table}[h]
 \begin{tabular}{|c|c|c|c|c|c|c|c|}
\hline
\backslashbox{Entropy}{Method}& Exact&LSNRD1&LSNRD2&LSST1&LSST2&Aquila1&Aquila2\\
\hline
$S_{AB}^X$&2.126&2.241&2.198&1.527&1.558&2.944&2.874\\
\hline
$S_{AB}^X ({\rm Trunc.})$&1.6476&1.734&1.740&1.504&1.527&2.035&1.901\\
\hline
$S_{A}^X$&1.337&1.374&1.354&1.117&1.134&1.685&1.681\\
\hline
$S_{A}^X({\rm Trunc.})$&1.131&1.149&1.167&1.115&1.108&1.288&1.233\\
\hline
$1.25(2S_A^X-S_{AB}^X)$&0.686&0.634&0.639&0.883&0.888&0.531&0.611\\
\hline
$1.25(2S_A^X-S_{AB}^X)({\rm Trunc.})$&0.769&0.706&0.742&0.907&0.861&0.676&0.707\\
\hline
\end{tabular}
\caption{\label{tab:ent} Entropies calculated with the full data set and truncated data sets (Trunc.) where observations with less than 10 events are discarded.}
\end{table}

\newpage
 \subsection{Entropies for a 5 rung ladder and $\Delta/\Omega=3.5$}
 \label{app3}
We repeated the calculation for ladder-shaped arrays with 5 rungs and 2 legs and different aspect ratios. We found very similar results for the same approximate constant 1.25 as shown in Fig. \ref{fig:ladder}. Fig. \ref{fig:ladderbound} makes clear that the inequality \beq
\label{eq:bound}
S_{A}^{vN}\geq 2S_A^X-S_{AB}^X,
\enq
 holds.

\begin{figure}[h]
\includegraphics[width=6.cm]{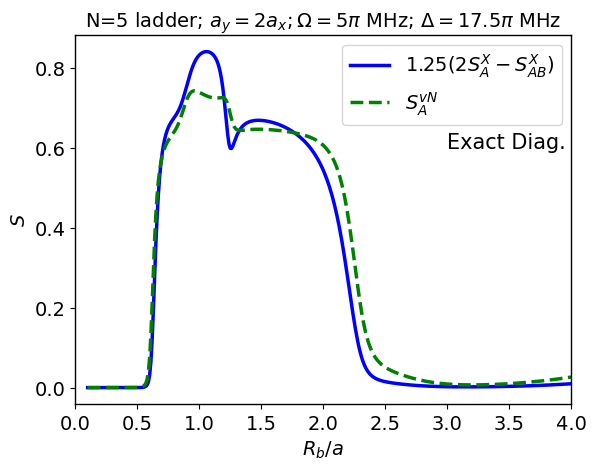} 
\includegraphics[width=6.cm]{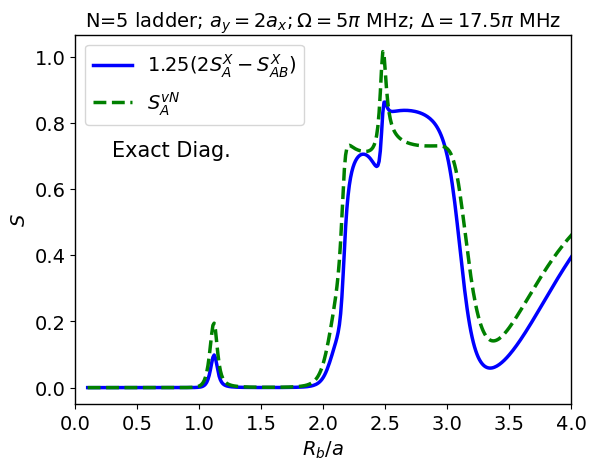} 
\caption{\label{fig:ladder} $1.25(2S_A^X-S_{AB}^X)$ and $S^{vN}
_A$ as a function of $R_b/a_x$ for $a_y=0.5a_x$ (left) and $a_y=2a_x$ (right).}
\end{figure}
\begin{figure}[h]
\includegraphics[width=6.cm]{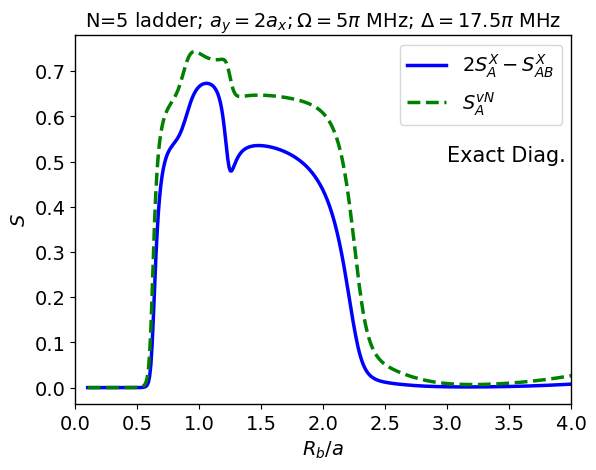} 
\includegraphics[width=6.cm]{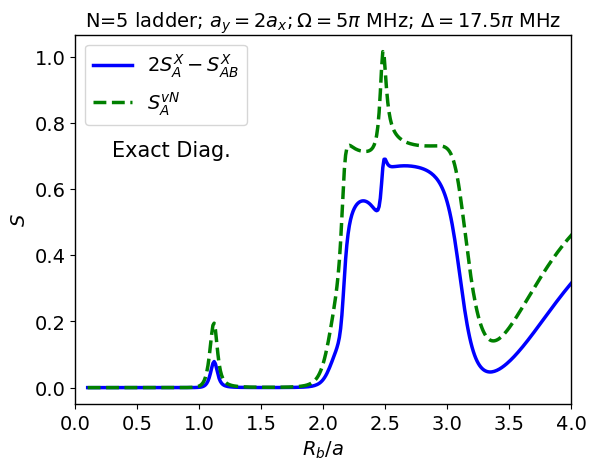} 
\caption{\label{fig:ladderbound} $2S_A^X-S_{AB}^X$ and $S^{vN}
_A$ as a function of $R_b/a_x$ for $a_y=0.5a_x$ (left) and $a_y=2a_x$ (right).}
\end{figure}

\vfill
\newpage
\subsection{Verification of the bound for two qubits}
\label{2q}

In this section, we consider a general two-qubit state
\beq
\ket{\psi}=\sum _{n_1 n_2=0,1} c _{n_1 n_2}\ket{n_1 n_2}.
\enq
The reduced two by two density matrix can be calculated exactly and we obtain 
\beq
S_A^{vN}=-\sum_{\pm} \lambda _{\pm} \ln(\lambda _{\pm}),
\enq
with
\beq 
\lambda _{\pm}=\frac{1}{2}(1\pm \sqrt{1-4 {\rm Det}}),
\enq
with the determinant 
\beq
{\rm Det}=p_{01}p_{10}+p_{00} p_{11}-2\sqrt{p_{00} p_{11}p_{01}p_{10}} \cos(\xi),
\enq
and a complex phase
\beq
\exp(i\xi)=(c_{00} c_{11}c_{01}^\star c_{10}^\star)/|c_{00} c_{11}c_{01} c_{10}|.
\enq
Det increases when $\xi$ increases from 0 to $\pi$. 
$S_A^{vN}=0$ when Det=0 and $\ln2$ when Det=1/4 and in general increases with Det. 
Consequently, we have the inequality
\beq
S_A^{vN}(p_{\{n\}},\xi)\geq S_A^{vN}(p_{\{n\}},\xi=0).
\enq
The inequality
\beq
S^{vN}(|c_{n_1n_2}|,\xi=0)\geq 2S_A^X-S_{AB}^X.
\enq
is verified numerically in Fig. \ref{fig:bounds}  using spherical coordinates to parametrize the norms of the coefficients expressed in spherical coordinates $|c_{00}| = \cos \phi \sin \theta;
|c_{11}| = \sin \phi \sin \theta;
|c_{01}|=|c_{10}| = \cos\theta/\sqrt{2}$ with $0\leq \theta,\phi \leq \pi/2$. This enforces the constraint that the probabilities add to one.
\begin{figure}[h]
\includegraphics[width=7.cm]{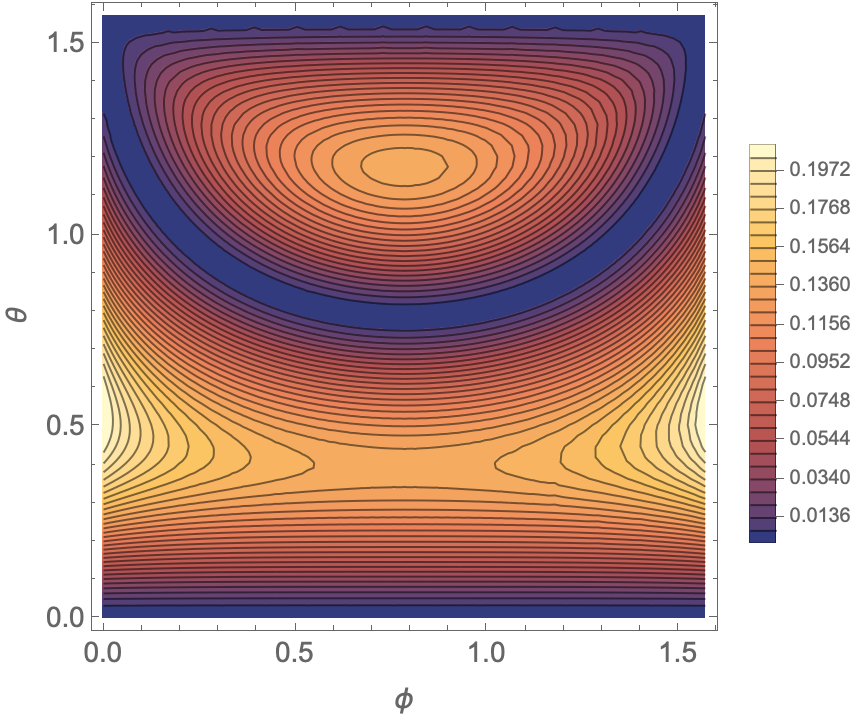} 
\caption{\label{fig:bounds} $S^{vN}(\xi=0)-(2S_A^X-S_{AB}^X)$ for  two qubits using the spherical coordinates described in the text.}
\end{figure}
\newpage

\hfill
\break
\subsection{Other bitstring proxies}
The adapted bitstring entropies  for the second order R\'enyi entropies can be defined as 
\beq
S_{2AB}^{ X}\equiv -\ln \sum_{\{n\}} p_{\{n\}}^2, \ {\rm and}\ S_{2A}^{X}\equiv -\ln \sum_{\{n\}_A} p_{\{n\}_A}^2. 
\enq
We can then compare $2S_{2A}^X-S_{2AB}^X$ with 
the quantum second order R\'enyi entropy $S_{2A}^{R}=-\ln({\rm Tr} \rho_A^2)$.
The numerical results for a 10 atom chain using exact diagonalization are displayed in Fig. \ref{fig:3echainR}.
This provides support for 
\beq
\label{eq:mainrenyi}
2S_{2A}^X-S_{2AB}^X\approx S_{2A}^{R} .
\enq
However, it appears that $2S_{2A}^X-S_{2AB}^X$ is not always positive.
\begin{figure}[h]
\includegraphics[width=7.cm]{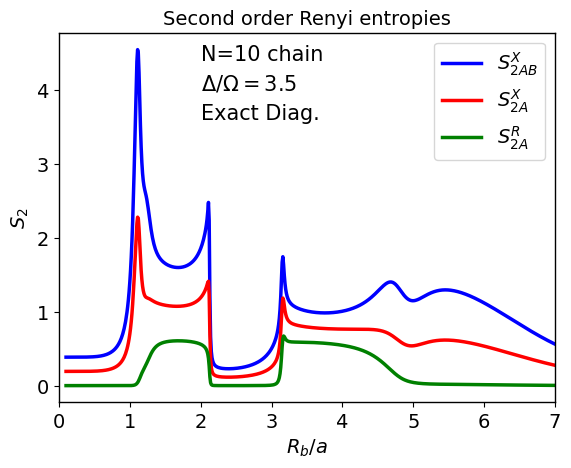} 
\includegraphics[width=7.2cm]{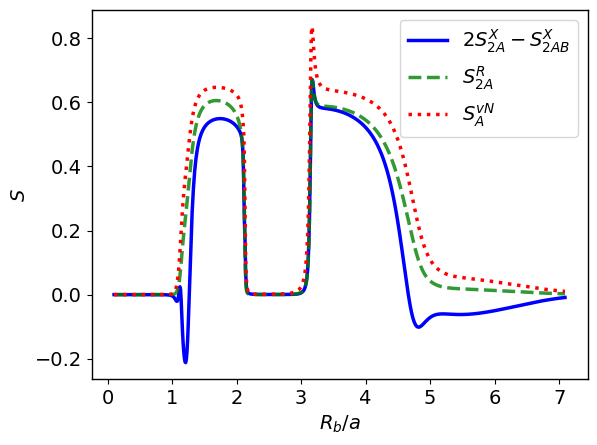} 
\caption{\label{fig:3echainR} Second order R\'enyi entropies.  Left: $S_{2AB}^X$ (top), $S_{2A}^X$  (middle) and $S_{2A}^{R}$(bottom) ; Right: $2S_{2A}^X-S_{2AB}^X$, $S_{2A}^{R}$, and    $S_A^{vN}$  for a chain of 10 atoms with $\Delta/\Omega=3.5$ as a function of $R_b/a_x$.}
\end{figure}

We also considered the adapted configurational entropy
\beq
 C \equiv 
\sum_{\{n\}} |p_{\{n\}}-p_{\{n\}_A}p_{\{n\}_B}|,
\enq
and the modified version introduced in the main text:
 \beq
 LC \equiv \ln(1+
\sum_{\{n\}} |p_{\{n\}}-p_{\{n\}_A}p_{\{n\}_B}|).
\enq
This quantity provides good estimates of $S_A^{vN}$ but is not always a lower bound.
%depends on the $p_{\{n\}}$ only and has similar limitations. Comparisons of the different approaches are under %study.
\begin{figure}[h]
\includegraphics[width=7.cm]{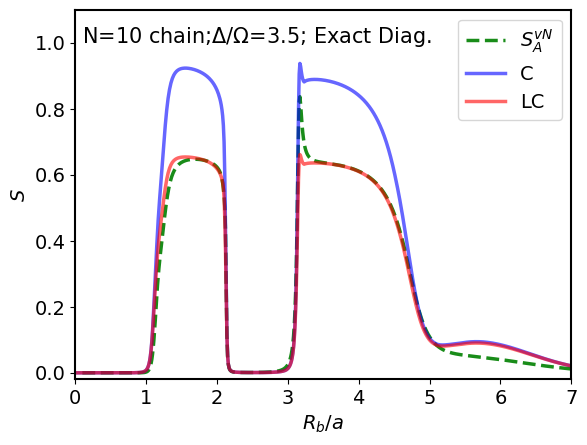} 
\caption{$C$, $LC$, and $S_A^{vN}$  for a chain of 10 atoms with $\Delta/\Omega=3.5$ as a function of $R_b/a_x$.}
\end{figure}
\end{widetext}
\newpage
$\  $
\newpage
$\ $
\newpage
%\bibliography{bibli}

\begin{thebibliography}{60}%
\makeatletter
\providecommand \@ifxundefined [1]{%
 \@ifx{#1\undefined}
}%
\providecommand \@ifnum [1]{%
 \ifnum #1\expandafter \@firstoftwo
 \else \expandafter \@secondoftwo
 \fi
}%
\providecommand \@ifx [1]{%
 \ifx #1\expandafter \@firstoftwo
 \else \expandafter \@secondoftwo
 \fi
}%
\providecommand \natexlab [1]{#1}%
\providecommand \enquote  [1]{``#1''}%
\providecommand \bibnamefont  [1]{#1}%
\providecommand \bibfnamefont [1]{#1}%
\providecommand \citenamefont [1]{#1}%
\providecommand \href@noop [0]{\@secondoftwo}%
\providecommand \href [0]{\begingroup \@sanitize@url \@href}%
\providecommand \@href[1]{\@@startlink{#1}\@@href}%
\providecommand \@@href[1]{\endgroup#1\@@endlink}%
\providecommand \@sanitize@url [0]{\catcode `\\12\catcode `\$12\catcode
  `\&12\catcode `\#12\catcode `\^12\catcode `\_12\catcode `\%12\relax}%
\providecommand \@@startlink[1]{}%
\providecommand \@@endlink[0]{}%
\providecommand \url  [0]{\begingroup\@sanitize@url \@url }%
\providecommand \@url [1]{\endgroup\@href {#1}{\urlprefix }}%
\providecommand \urlprefix  [0]{URL }%
\providecommand \Eprint [0]{\href }%
\providecommand \doibase [0]{https://doi.org/}%
\providecommand \selectlanguage [0]{\@gobble}%
\providecommand \bibinfo  [0]{\@secondoftwo}%
\providecommand \bibfield  [0]{\@secondoftwo}%
\providecommand \translation [1]{[#1]}%
\providecommand \BibitemOpen [0]{}%
\providecommand \bibitemStop [0]{}%
\providecommand \bibitemNoStop [0]{.\EOS\space}%
\providecommand \EOS [0]{\spacefactor3000\relax}%
\providecommand \BibitemShut  [1]{\csname bibitem#1\endcsname}%
\let\auto@bib@innerbib\@empty
%</preamble>
\bibitem [{\citenamefont {Amico}\ \emph {et~al.}(2008)\citenamefont {Amico},
  \citenamefont {Fazio}, \citenamefont {Osterloh},\ and\ \citenamefont
  {Vedral}}]{amico2008entanglement}%
  \BibitemOpen
  \bibfield  {author} {\bibinfo {author} {\bibfnamefont {L.}~\bibnamefont
  {Amico}}, \bibinfo {author} {\bibfnamefont {R.}~\bibnamefont {Fazio}},
  \bibinfo {author} {\bibfnamefont {A.}~\bibnamefont {Osterloh}},\ and\
  \bibinfo {author} {\bibfnamefont {V.}~\bibnamefont {Vedral}},\ }\bibfield
  {title} {\bibinfo {title} {Entanglement in many-body systems},\ }\href@noop
  {} {\bibfield  {journal} {\bibinfo  {journal} {Reviews of Modern Physics}\
  }\textbf {\bibinfo {volume} {80}},\ \bibinfo {pages} {517} (\bibinfo {year}
  {2008})}\BibitemShut {NoStop}%
\bibitem [{\citenamefont {Eisert}\ \emph {et~al.}(2010)\citenamefont {Eisert},
  \citenamefont {Cramer},\ and\ \citenamefont {Plenio}}]{Eisert:2008ur}%
  \BibitemOpen
  \bibfield  {author} {\bibinfo {author} {\bibfnamefont {J.}~\bibnamefont
  {Eisert}}, \bibinfo {author} {\bibfnamefont {M.}~\bibnamefont {Cramer}},\
  and\ \bibinfo {author} {\bibfnamefont {M.~B.}\ \bibnamefont {Plenio}},\
  }\bibfield  {title} {\bibinfo {title} {{Area laws for the entanglement
  entropy - a review}},\ }\href {https://doi.org/10.1103/RevModPhys.82.277}
  {\bibfield  {journal} {\bibinfo  {journal} {Rev. Mod. Phys.}\ }\textbf
  {\bibinfo {volume} {82}},\ \bibinfo {pages} {277} (\bibinfo {year} {2010})},\
  \Eprint {https://arxiv.org/abs/0808.3773} {arXiv:0808.3773 [quant-ph]}
  \BibitemShut {NoStop}%
\bibitem [{\citenamefont {Abanin}\ \emph {et~al.}(2019)\citenamefont {Abanin},
  \citenamefont {Altman}, \citenamefont {Bloch},\ and\ \citenamefont
  {Serbyn}}]{Abanin_2019}%
  \BibitemOpen
  \bibfield  {author} {\bibinfo {author} {\bibfnamefont {D.~A.}\ \bibnamefont
  {Abanin}}, \bibinfo {author} {\bibfnamefont {E.}~\bibnamefont {Altman}},
  \bibinfo {author} {\bibfnamefont {I.}~\bibnamefont {Bloch}},\ and\ \bibinfo
  {author} {\bibfnamefont {M.}~\bibnamefont {Serbyn}},\ }\bibfield  {title}
  {\bibinfo {title} {Colloquium : Many-body localization, thermalization, and
  entanglement},\ }\bibfield  {journal} {\bibinfo  {journal} {Reviews of Modern
  Physics}\ }\textbf {\bibinfo {volume} {91}},\ \href
  {https://doi.org/10.1103/revmodphys.91.021001} {10.1103/revmodphys.91.021001}
  (\bibinfo {year} {2019})\BibitemShut {NoStop}%
\bibitem [{\citenamefont {Cirac}\ \emph {et~al.}(2021)\citenamefont {Cirac},
  \citenamefont {Perez-Garcia}, \citenamefont {Schuch},\ and\ \citenamefont
  {Verstraete}}]{Cirac:2020obd}%
  \BibitemOpen
  \bibfield  {author} {\bibinfo {author} {\bibfnamefont {J.~I.}\ \bibnamefont
  {Cirac}}, \bibinfo {author} {\bibfnamefont {D.}~\bibnamefont {Perez-Garcia}},
  \bibinfo {author} {\bibfnamefont {N.}~\bibnamefont {Schuch}},\ and\ \bibinfo
  {author} {\bibfnamefont {F.}~\bibnamefont {Verstraete}},\ }\bibfield  {title}
  {\bibinfo {title} {{Matrix product states and projected entangled pair
  states: Concepts, symmetries, theorems}},\ }\href
  {https://doi.org/10.1103/RevModPhys.93.045003} {\bibfield  {journal}
  {\bibinfo  {journal} {Rev. Mod. Phys.}\ }\textbf {\bibinfo {volume} {93}},\
  \bibinfo {pages} {045003} (\bibinfo {year} {2021})},\ \Eprint
  {https://arxiv.org/abs/2011.12127} {arXiv:2011.12127 [quant-ph]} \BibitemShut
  {NoStop}%
\bibitem [{\citenamefont {Ghosh}\ \emph {et~al.}(2015)\citenamefont {Ghosh},
  \citenamefont {Soni},\ and\ \citenamefont {Trivedi}}]{Ghosh:2015iwa}%
  \BibitemOpen
  \bibfield  {author} {\bibinfo {author} {\bibfnamefont {S.}~\bibnamefont
  {Ghosh}}, \bibinfo {author} {\bibfnamefont {R.~M.}\ \bibnamefont {Soni}},\
  and\ \bibinfo {author} {\bibfnamefont {S.~P.}\ \bibnamefont {Trivedi}},\
  }\bibfield  {title} {\bibinfo {title} {{On The Entanglement Entropy For Gauge
  Theories}},\ }\href {https://doi.org/10.1007/JHEP09(2015)069} {\bibfield
  {journal} {\bibinfo  {journal} {JHEP}\ }\textbf {\bibinfo {volume} {09}},\
  \bibinfo {pages} {069}},\ \Eprint {https://arxiv.org/abs/1501.02593}
  {arXiv:1501.02593 [hep-th]} \BibitemShut {NoStop}%
\bibitem [{\citenamefont {Van~Acoleyen}\ \emph {et~al.}(2016)\citenamefont
  {Van~Acoleyen}, \citenamefont {Bultinck}, \citenamefont {Haegeman},
  \citenamefont {Marien}, \citenamefont {Scholz},\ and\ \citenamefont
  {Verstraete}}]{VanAcoleyen:2015ccp}%
  \BibitemOpen
  \bibfield  {author} {\bibinfo {author} {\bibfnamefont {K.}~\bibnamefont
  {Van~Acoleyen}}, \bibinfo {author} {\bibfnamefont {N.}~\bibnamefont
  {Bultinck}}, \bibinfo {author} {\bibfnamefont {J.}~\bibnamefont {Haegeman}},
  \bibinfo {author} {\bibfnamefont {M.}~\bibnamefont {Marien}}, \bibinfo
  {author} {\bibfnamefont {V.~B.}\ \bibnamefont {Scholz}},\ and\ \bibinfo
  {author} {\bibfnamefont {F.}~\bibnamefont {Verstraete}},\ }\bibfield  {title}
  {\bibinfo {title} {{The entanglement of distillation for gauge theories}},\
  }\href {https://doi.org/10.1103/PhysRevLett.117.131602} {\bibfield  {journal}
  {\bibinfo  {journal} {Phys. Rev. Lett.}\ }\textbf {\bibinfo {volume} {117}},\
  \bibinfo {pages} {131602} (\bibinfo {year} {2016})},\ \Eprint
  {https://arxiv.org/abs/1511.04369} {arXiv:1511.04369 [quant-ph]} \BibitemShut
  {NoStop}%
\bibitem [{\citenamefont {Ba\~nuls}\ \emph {et~al.}(2017)\citenamefont
  {Ba\~nuls}, \citenamefont {Cichy}, \citenamefont {Cirac}, \citenamefont
  {Jansen},\ and\ \citenamefont {K\"uhn}}]{Banuls:2017ena}%
  \BibitemOpen
  \bibfield  {author} {\bibinfo {author} {\bibfnamefont {M.~C.}\ \bibnamefont
  {Ba\~nuls}}, \bibinfo {author} {\bibfnamefont {K.}~\bibnamefont {Cichy}},
  \bibinfo {author} {\bibfnamefont {J.~I.}\ \bibnamefont {Cirac}}, \bibinfo
  {author} {\bibfnamefont {K.}~\bibnamefont {Jansen}},\ and\ \bibinfo {author}
  {\bibfnamefont {S.}~\bibnamefont {K\"uhn}},\ }\bibfield  {title} {\bibinfo
  {title} {{Efficient basis formulation for 1+1 dimensional SU(2) lattice gauge
  theory: Spectral calculations with matrix product states}},\ }\href
  {https://doi.org/10.1103/PhysRevX.7.041046} {\bibfield  {journal} {\bibinfo
  {journal} {Phys. Rev. X}\ }\textbf {\bibinfo {volume} {7}},\ \bibinfo {pages}
  {041046} (\bibinfo {year} {2017})},\ \Eprint
  {https://arxiv.org/abs/1707.06434} {arXiv:1707.06434 [hep-lat]} \BibitemShut
  {NoStop}%
\bibitem [{\citenamefont {Knaute}\ \emph {et~al.}(2024)\citenamefont {Knaute},
  \citenamefont {Feuerstein},\ and\ \citenamefont {Zohar}}]{Knaute:2024wfh}%
  \BibitemOpen
  \bibfield  {author} {\bibinfo {author} {\bibfnamefont {J.}~\bibnamefont
  {Knaute}}, \bibinfo {author} {\bibfnamefont {M.}~\bibnamefont {Feuerstein}},\
  and\ \bibinfo {author} {\bibfnamefont {E.}~\bibnamefont {Zohar}},\ }\bibfield
   {title} {\bibinfo {title} {{Entanglement and confinement in lattice gauge
  theory tensor networks}},\ }\href {https://doi.org/10.1007/JHEP02(2024)174}
  {\bibfield  {journal} {\bibinfo  {journal} {JHEP}\ }\textbf {\bibinfo
  {volume} {02}},\ \bibinfo {pages} {174}},\ \Eprint
  {https://arxiv.org/abs/2401.01930} {arXiv:2401.01930 [quant-ph]} \BibitemShut
  {NoStop}%
\bibitem [{\citenamefont {Kharzeev}\ and\ \citenamefont
  {Levin}(2017)}]{Kharzeev:2017qzs}%
  \BibitemOpen
  \bibfield  {author} {\bibinfo {author} {\bibfnamefont {D.~E.}\ \bibnamefont
  {Kharzeev}}\ and\ \bibinfo {author} {\bibfnamefont {E.~M.}\ \bibnamefont
  {Levin}},\ }\bibfield  {title} {\bibinfo {title} {{Deep inelastic scattering
  as a probe of entanglement}},\ }\href
  {https://doi.org/10.1103/PhysRevD.95.114008} {\bibfield  {journal} {\bibinfo
  {journal} {Phys. Rev. D}\ }\textbf {\bibinfo {volume} {95}},\ \bibinfo
  {pages} {114008} (\bibinfo {year} {2017})},\ \Eprint
  {https://arxiv.org/abs/1702.03489} {arXiv:1702.03489 [hep-ph]} \BibitemShut
  {NoStop}%
\bibitem [{\citenamefont {Baker}\ and\ \citenamefont
  {Kharzeev}(2018)}]{PhysRevD.98.054007}%
  \BibitemOpen
  \bibfield  {author} {\bibinfo {author} {\bibfnamefont {O.~K.}\ \bibnamefont
  {Baker}}\ and\ \bibinfo {author} {\bibfnamefont {D.~E.}\ \bibnamefont
  {Kharzeev}},\ }\bibfield  {title} {\bibinfo {title} {Thermal radiation and
  entanglement in proton-proton collisions at energies available at the cern
  large hadron collider},\ }\href {https://doi.org/10.1103/PhysRevD.98.054007}
  {\bibfield  {journal} {\bibinfo  {journal} {Phys. Rev. D}\ }\textbf {\bibinfo
  {volume} {98}},\ \bibinfo {pages} {054007} (\bibinfo {year}
  {2018})}\BibitemShut {NoStop}%
\bibitem [{\citenamefont {Zhang}\ \emph {et~al.}(2022)\citenamefont {Zhang},
  \citenamefont {Hao}, \citenamefont {Kharzeev},\ and\ \citenamefont
  {Korepin}}]{Zhang:2021hra}%
  \BibitemOpen
  \bibfield  {author} {\bibinfo {author} {\bibfnamefont {K.}~\bibnamefont
  {Zhang}}, \bibinfo {author} {\bibfnamefont {K.}~\bibnamefont {Hao}}, \bibinfo
  {author} {\bibfnamefont {D.}~\bibnamefont {Kharzeev}},\ and\ \bibinfo
  {author} {\bibfnamefont {V.}~\bibnamefont {Korepin}},\ }\bibfield  {title}
  {\bibinfo {title} {{Entanglement entropy production in deep inelastic
  scattering}},\ }\href {https://doi.org/10.1103/PhysRevD.105.014002}
  {\bibfield  {journal} {\bibinfo  {journal} {Phys. Rev. D}\ }\textbf {\bibinfo
  {volume} {105}},\ \bibinfo {pages} {014002} (\bibinfo {year} {2022})},\
  \Eprint {https://arxiv.org/abs/2110.04881} {arXiv:2110.04881 [quant-ph]}
  \BibitemShut {NoStop}%
\bibitem [{\citenamefont {Beane}\ \emph {et~al.}(2019)\citenamefont {Beane},
  \citenamefont {Kaplan}, \citenamefont {Klco},\ and\ \citenamefont
  {Savage}}]{Beane:2018oxh}%
  \BibitemOpen
  \bibfield  {author} {\bibinfo {author} {\bibfnamefont {S.~R.}\ \bibnamefont
  {Beane}}, \bibinfo {author} {\bibfnamefont {D.~B.}\ \bibnamefont {Kaplan}},
  \bibinfo {author} {\bibfnamefont {N.}~\bibnamefont {Klco}},\ and\ \bibinfo
  {author} {\bibfnamefont {M.~J.}\ \bibnamefont {Savage}},\ }\bibfield  {title}
  {\bibinfo {title} {{Entanglement Suppression and Emergent Symmetries of
  Strong Interactions}},\ }\href
  {https://doi.org/10.1103/PhysRevLett.122.102001} {\bibfield  {journal}
  {\bibinfo  {journal} {Phys. Rev. Lett.}\ }\textbf {\bibinfo {volume} {122}},\
  \bibinfo {pages} {102001} (\bibinfo {year} {2019})},\ \Eprint
  {https://arxiv.org/abs/1812.03138} {arXiv:1812.03138 [nucl-th]} \BibitemShut
  {NoStop}%
\bibitem [{\citenamefont {Robin}\ \emph {et~al.}(2021)\citenamefont {Robin},
  \citenamefont {Savage},\ and\ \citenamefont {Pillet}}]{Robin:2020aeh}%
  \BibitemOpen
  \bibfield  {author} {\bibinfo {author} {\bibfnamefont {C.}~\bibnamefont
  {Robin}}, \bibinfo {author} {\bibfnamefont {M.~J.}\ \bibnamefont {Savage}},\
  and\ \bibinfo {author} {\bibfnamefont {N.}~\bibnamefont {Pillet}},\
  }\bibfield  {title} {\bibinfo {title} {{Entanglement Rearrangement in
  Self-Consistent Nuclear Structure Calculations}},\ }\href
  {https://doi.org/10.1103/PhysRevC.103.034325} {\bibfield  {journal} {\bibinfo
   {journal} {Phys. Rev. C}\ }\textbf {\bibinfo {volume} {103}},\ \bibinfo
  {pages} {034325} (\bibinfo {year} {2021})},\ \Eprint
  {https://arxiv.org/abs/2007.09157} {arXiv:2007.09157 [nucl-th]} \BibitemShut
  {NoStop}%
\bibitem [{\citenamefont {Vidal}\ \emph {et~al.}(2003)\citenamefont {Vidal},
  \citenamefont {Latorre}, \citenamefont {Rico},\ and\ \citenamefont
  {Kitaev}}]{PhysRevLett.90.227902}%
  \BibitemOpen
  \bibfield  {author} {\bibinfo {author} {\bibfnamefont {G.}~\bibnamefont
  {Vidal}}, \bibinfo {author} {\bibfnamefont {J.~I.}\ \bibnamefont {Latorre}},
  \bibinfo {author} {\bibfnamefont {E.}~\bibnamefont {Rico}},\ and\ \bibinfo
  {author} {\bibfnamefont {A.}~\bibnamefont {Kitaev}},\ }\bibfield  {title}
  {\bibinfo {title} {Entanglement in quantum critical phenomena},\ }\href
  {https://doi.org/10.1103/PhysRevLett.90.227902} {\bibfield  {journal}
  {\bibinfo  {journal} {Phys. Rev. Lett.}\ }\textbf {\bibinfo {volume} {90}},\
  \bibinfo {pages} {227902} (\bibinfo {year} {2003})}\BibitemShut {NoStop}%
\bibitem [{\citenamefont {Korepin}(2004)}]{PhysRevLett.92.096402}%
  \BibitemOpen
  \bibfield  {author} {\bibinfo {author} {\bibfnamefont {V.~E.}\ \bibnamefont
  {Korepin}},\ }\bibfield  {title} {\bibinfo {title} {Universality of entropy
  scaling in one dimensional gapless models},\ }\href
  {https://doi.org/10.1103/PhysRevLett.92.096402} {\bibfield  {journal}
  {\bibinfo  {journal} {Phys. Rev. Lett.}\ }\textbf {\bibinfo {volume} {92}},\
  \bibinfo {pages} {096402} (\bibinfo {year} {2004})}\BibitemShut {NoStop}%
\bibitem [{\citenamefont {Calabrese}\ and\ \citenamefont
  {Cardy}(2006)}]{Calabrese:2005zw}%
  \BibitemOpen
  \bibfield  {author} {\bibinfo {author} {\bibfnamefont {P.}~\bibnamefont
  {Calabrese}}\ and\ \bibinfo {author} {\bibfnamefont {J.~L.}\ \bibnamefont
  {Cardy}},\ }\bibfield  {title} {\bibinfo {title} {{Entanglement entropy and
  quantum field theory: A Non-technical introduction}},\ }\href
  {https://doi.org/10.1142/S021974990600192X} {\bibfield  {journal} {\bibinfo
  {journal} {Int. J. Quant. Inf.}\ }\textbf {\bibinfo {volume} {4}},\ \bibinfo
  {pages} {429} (\bibinfo {year} {2006})},\ \Eprint
  {https://arxiv.org/abs/quant-ph/0505193} {arXiv:quant-ph/0505193}
  \BibitemShut {NoStop}%
\bibitem [{\citenamefont {Ryu}\ and\ \citenamefont
  {Takayanagi}(2006)}]{Ryu:2006bv}%
  \BibitemOpen
  \bibfield  {author} {\bibinfo {author} {\bibfnamefont {S.}~\bibnamefont
  {Ryu}}\ and\ \bibinfo {author} {\bibfnamefont {T.}~\bibnamefont
  {Takayanagi}},\ }\bibfield  {title} {\bibinfo {title} {{Holographic
  derivation of entanglement entropy from AdS/CFT}},\ }\href
  {https://doi.org/10.1103/PhysRevLett.96.181602} {\bibfield  {journal}
  {\bibinfo  {journal} {Phys. Rev. Lett.}\ }\textbf {\bibinfo {volume} {96}},\
  \bibinfo {pages} {181602} (\bibinfo {year} {2006})},\ \Eprint
  {https://arxiv.org/abs/hep-th/0603001} {arXiv:hep-th/0603001} \BibitemShut
  {NoStop}%
\bibitem [{\citenamefont {Abanin}\ and\ \citenamefont
  {Demler}(2012)}]{PhysRevLett.109.020504}%
  \BibitemOpen
  \bibfield  {author} {\bibinfo {author} {\bibfnamefont {D.~A.}\ \bibnamefont
  {Abanin}}\ and\ \bibinfo {author} {\bibfnamefont {E.}~\bibnamefont
  {Demler}},\ }\bibfield  {title} {\bibinfo {title} {Measuring entanglement
  entropy of a generic many-body system with a quantum switch},\ }\href
  {https://doi.org/10.1103/PhysRevLett.109.020504} {\bibfield  {journal}
  {\bibinfo  {journal} {Phys. Rev. Lett.}\ }\textbf {\bibinfo {volume} {109}},\
  \bibinfo {pages} {020504} (\bibinfo {year} {2012})}\BibitemShut {NoStop}%
\bibitem [{\citenamefont {Daley}\ \emph {et~al.}(2012)\citenamefont {Daley},
  \citenamefont {Pichler}, \citenamefont {Schachenmayer},\ and\ \citenamefont
  {Zoller}}]{PhysRevLett.109.020505}%
  \BibitemOpen
  \bibfield  {author} {\bibinfo {author} {\bibfnamefont {A.~J.}\ \bibnamefont
  {Daley}}, \bibinfo {author} {\bibfnamefont {H.}~\bibnamefont {Pichler}},
  \bibinfo {author} {\bibfnamefont {J.}~\bibnamefont {Schachenmayer}},\ and\
  \bibinfo {author} {\bibfnamefont {P.}~\bibnamefont {Zoller}},\ }\bibfield
  {title} {\bibinfo {title} {Measuring entanglement growth in quench dynamics
  of bosons in an optical lattice},\ }\href
  {https://doi.org/10.1103/PhysRevLett.109.020505} {\bibfield  {journal}
  {\bibinfo  {journal} {Phys. Rev. Lett.}\ }\textbf {\bibinfo {volume} {109}},\
  \bibinfo {pages} {020505} (\bibinfo {year} {2012})}\BibitemShut {NoStop}%
\bibitem [{\citenamefont {Islam}\ \emph {et~al.}(2015)\citenamefont {Islam},
  \citenamefont {Ma}, \citenamefont {Preiss}, \citenamefont {Eric~Tai},
  \citenamefont {Lukin}, \citenamefont {Rispoli},\ and\ \citenamefont
  {Greiner}}]{Islam:2015mom}%
  \BibitemOpen
  \bibfield  {author} {\bibinfo {author} {\bibfnamefont {R.}~\bibnamefont
  {Islam}}, \bibinfo {author} {\bibfnamefont {R.}~\bibnamefont {Ma}}, \bibinfo
  {author} {\bibfnamefont {P.~M.}\ \bibnamefont {Preiss}}, \bibinfo {author}
  {\bibfnamefont {M.}~\bibnamefont {Eric~Tai}}, \bibinfo {author}
  {\bibfnamefont {A.}~\bibnamefont {Lukin}}, \bibinfo {author} {\bibfnamefont
  {M.}~\bibnamefont {Rispoli}},\ and\ \bibinfo {author} {\bibfnamefont
  {M.}~\bibnamefont {Greiner}},\ }\bibfield  {title} {\bibinfo {title}
  {Measuring entanglement entropy in a quantum many-body system},\ }\href
  {https://doi.org/10.1038/nature15750} {\bibfield  {journal} {\bibinfo
  {journal} {Nature}\ }\textbf {\bibinfo {volume} {528}},\ \bibinfo {pages}
  {77–83} (\bibinfo {year} {2015})}\BibitemShut {NoStop}%
\bibitem [{\citenamefont {Kaufman}\ \emph {et~al.}(2016)\citenamefont
  {Kaufman}, \citenamefont {Tai}, \citenamefont {Lukin}, \citenamefont
  {Rispoli}, \citenamefont {Schittko}, \citenamefont {Preiss},\ and\
  \citenamefont {Greiner}}]{Kaufman:2016mif}%
  \BibitemOpen
  \bibfield  {author} {\bibinfo {author} {\bibfnamefont {A.~M.}\ \bibnamefont
  {Kaufman}}, \bibinfo {author} {\bibfnamefont {M.~E.}\ \bibnamefont {Tai}},
  \bibinfo {author} {\bibfnamefont {A.}~\bibnamefont {Lukin}}, \bibinfo
  {author} {\bibfnamefont {M.}~\bibnamefont {Rispoli}}, \bibinfo {author}
  {\bibfnamefont {R.}~\bibnamefont {Schittko}}, \bibinfo {author}
  {\bibfnamefont {P.~M.}\ \bibnamefont {Preiss}},\ and\ \bibinfo {author}
  {\bibfnamefont {M.}~\bibnamefont {Greiner}},\ }\bibfield  {title} {\bibinfo
  {title} {Quantum thermalization through entanglement in an isolated many-body
  system},\ }\href {https://doi.org/10.1126/science.aaf6725} {\bibfield
  {journal} {\bibinfo  {journal} {Science}\ }\textbf {\bibinfo {volume}
  {353}},\ \bibinfo {pages} {794–800} (\bibinfo {year} {2016})}\BibitemShut
  {NoStop}%
\bibitem [{\citenamefont {Unmuth-Yockey}\ \emph {et~al.}(2017)\citenamefont
  {Unmuth-Yockey}, \citenamefont {Zhang}, \citenamefont {Preiss}, \citenamefont
  {Yang}, \citenamefont {Tsai},\ and\ \citenamefont
  {Meurice}}]{Unmuth-Yockey:2016znu}%
  \BibitemOpen
  \bibfield  {author} {\bibinfo {author} {\bibfnamefont {J.}~\bibnamefont
  {Unmuth-Yockey}}, \bibinfo {author} {\bibfnamefont {J.}~\bibnamefont
  {Zhang}}, \bibinfo {author} {\bibfnamefont {P.~M.}\ \bibnamefont {Preiss}},
  \bibinfo {author} {\bibfnamefont {L.-P.}\ \bibnamefont {Yang}}, \bibinfo
  {author} {\bibfnamefont {S.~W.}\ \bibnamefont {Tsai}},\ and\ \bibinfo
  {author} {\bibfnamefont {Y.}~\bibnamefont {Meurice}},\ }\bibfield  {title}
  {\bibinfo {title} {{Probing the conformal Calabrese-Cardy scaling with cold
  atoms}},\ }\href {https://doi.org/10.1103/PhysRevA.96.023603} {\bibfield
  {journal} {\bibinfo  {journal} {Phys. Rev. A}\ }\textbf {\bibinfo {volume}
  {96}},\ \bibinfo {pages} {023603} (\bibinfo {year} {2017})},\ \Eprint
  {https://arxiv.org/abs/1611.05016} {arXiv:1611.05016 [cond-mat.quant-gas]}
  \BibitemShut {NoStop}%
\bibitem [{\citenamefont {Shannon}\ and\ \citenamefont
  {Weaver}(1998)}]{shannon49}%
  \BibitemOpen
  \bibfield  {author} {\bibinfo {author} {\bibfnamefont {C.}~\bibnamefont
  {Shannon}}\ and\ \bibinfo {author} {\bibfnamefont {W.}~\bibnamefont
  {Weaver}},\ }\href {https://books.google.com/books?id=IZ77BwAAQBAJ} {\emph
  {\bibinfo {title} {The Mathematical Theory of Communication}}}\ (\bibinfo
  {publisher} {University of Illinois Press},\ \bibinfo {year}
  {1998})\BibitemShut {NoStop}%
\bibitem [{\citenamefont {Nielsen}\ and\ \citenamefont {Chuang}(2011)}]{nandc}%
  \BibitemOpen
  \bibfield  {author} {\bibinfo {author} {\bibfnamefont {M.~A.}\ \bibnamefont
  {Nielsen}}\ and\ \bibinfo {author} {\bibfnamefont {I.~L.}\ \bibnamefont
  {Chuang}},\ }\href
  {https://www.amazon.com/Quantum-Computation-Information-10th-Anniversary/dp/1107002176?SubscriptionId=AKIAIOBINVZYXZQZ2U3A&tag=chimbori05-20&linkCode=xm2&camp=2025&creative=165953&creativeASIN=1107002176}
  {\emph {\bibinfo {title} {Quantum Computation and Quantum Information: 10th
  Anniversary Edition}}}\ (\bibinfo  {publisher} {Cambridge University Press},\
  \bibinfo {year} {2011})\BibitemShut {NoStop}%
\bibitem [{\citenamefont {Preskill}(2022)}]{jpch10}%
  \BibitemOpen
  \bibfield  {author} {\bibinfo {author} {\bibfnamefont {J.}~\bibnamefont
  {Preskill}},\ }\href
  {https://www.preskill.caltech.edu/ph219/chap10_6A_2022.pdf} {\bibinfo {title}
  {Quantum information chapter 10. quantum shannon theory}} (\bibinfo {year}
  {2022})\BibitemShut {NoStop}%
\bibitem [{\citenamefont {Wurtz}\ \emph {et~al.}(2023)\citenamefont {Wurtz},
  \citenamefont {Bylinskii}, \citenamefont {Braverman}, \citenamefont
  {Amato-Grill}, \citenamefont {Cantu}, \citenamefont {Huber}, \citenamefont
  {Lukin}, \citenamefont {Liu}, \citenamefont {Weinberg}, \citenamefont {Long},
  \citenamefont {Wang}, \citenamefont {Gemelke},\ and\ \citenamefont
  {Keesling}}]{wurtz2023aquila}%
  \BibitemOpen
  \bibfield  {author} {\bibinfo {author} {\bibfnamefont {J.}~\bibnamefont
  {Wurtz}}, \bibinfo {author} {\bibfnamefont {A.}~\bibnamefont {Bylinskii}},
  \bibinfo {author} {\bibfnamefont {B.}~\bibnamefont {Braverman}}, \bibinfo
  {author} {\bibfnamefont {J.}~\bibnamefont {Amato-Grill}}, \bibinfo {author}
  {\bibfnamefont {S.~H.}\ \bibnamefont {Cantu}}, \bibinfo {author}
  {\bibfnamefont {F.}~\bibnamefont {Huber}}, \bibinfo {author} {\bibfnamefont
  {A.}~\bibnamefont {Lukin}}, \bibinfo {author} {\bibfnamefont
  {F.}~\bibnamefont {Liu}}, \bibinfo {author} {\bibfnamefont {P.}~\bibnamefont
  {Weinberg}}, \bibinfo {author} {\bibfnamefont {J.}~\bibnamefont {Long}},
  \bibinfo {author} {\bibfnamefont {S.-T.}\ \bibnamefont {Wang}}, \bibinfo
  {author} {\bibfnamefont {N.}~\bibnamefont {Gemelke}},\ and\ \bibinfo {author}
  {\bibfnamefont {A.}~\bibnamefont {Keesling}},\ }\href@noop {} {\bibinfo
  {title} {Aquila: Quera's 256-qubit neutral-atom quantum computer}} (\bibinfo
  {year} {2023}),\ \Eprint {https://arxiv.org/abs/2306.11727} {arXiv:2306.11727
  [quant-ph]} \BibitemShut {NoStop}%
\bibitem [{\citenamefont {Witten}(2020)}]{wittenmini}%
  \BibitemOpen
  \bibfield  {author} {\bibinfo {author} {\bibfnamefont {E.}~\bibnamefont
  {Witten}},\ }\bibfield  {title} {\bibinfo {title} {{A Mini-Introduction To
  Information Theory}},\ }\href {https://doi.org/10.1007/s40766-020-00004-5}
  {\bibfield  {journal} {\bibinfo  {journal} {Riv. Nuovo Cim.}\ }\textbf
  {\bibinfo {volume} {43}},\ \bibinfo {pages} {187} (\bibinfo {year} {2020})},\
  \Eprint {https://arxiv.org/abs/1805.11965} {arXiv:1805.11965 [hep-th]}
  \BibitemShut {NoStop}%
\bibitem [{\citenamefont {Preskill}\ and\ \citenamefont
  {Soleimanifar}(2024)}]{mehdi}%
  \BibitemOpen
  \bibfield  {author} {\bibinfo {author} {\bibfnamefont {J.}~\bibnamefont
  {Preskill}}\ and\ \bibinfo {author} {\bibfnamefont {M.}~\bibnamefont
  {Soleimanifar}},\ }\href@noop {} {\bibinfo {title} {Private communication}}
  (\bibinfo {year} {2024})\BibitemShut {NoStop}%
\bibitem [{\citenamefont {Kaufman}\ \emph {et~al.}(2024)\citenamefont
  {Kaufman}, \citenamefont {Corona}, \citenamefont {Ozzello}, \citenamefont
  {Asaduzzaman},\ and\ \citenamefont {Meurice}}]{avi}%
  \BibitemOpen
  \bibfield  {author} {\bibinfo {author} {\bibfnamefont {A.}~\bibnamefont
  {Kaufman}}, \bibinfo {author} {\bibfnamefont {J.}~\bibnamefont {Corona}},
  \bibinfo {author} {\bibfnamefont {Z.}~\bibnamefont {Ozzello}}, \bibinfo
  {author} {\bibfnamefont {M.}~\bibnamefont {Asaduzzaman}},\ and\ \bibinfo
  {author} {\bibfnamefont {Y.}~\bibnamefont {Meurice}},\ }\href@noop {}
  {\bibinfo {title} {{Improved entanglement entropy estimates from filtered
  bitstring probabilities}}} (\bibinfo {year} {2024}),\ \Eprint
  {https://arxiv.org/abs/2411.07092} {arXiv:2411.07092 [quant-ph]} \BibitemShut
  {NoStop}%
\bibitem [{\citenamefont {Ozzello}\ and\ \citenamefont
  {Meurice}(2024)}]{zaneprogress}%
  \BibitemOpen
  \bibfield  {author} {\bibinfo {author} {\bibfnamefont {Z.}~\bibnamefont
  {Ozzello}}\ and\ \bibinfo {author} {\bibfnamefont {Y.}~\bibnamefont
  {Meurice}},\ }\href@noop {} {\bibinfo {title} {Work in progress}} (\bibinfo
  {year} {2024})\BibitemShut {NoStop}%
\bibitem [{\citenamefont {Bernien}\ \emph {et~al.}(2017)\citenamefont
  {Bernien}, \citenamefont {Schwartz}, \citenamefont {Keesling}, \citenamefont
  {Levine}, \citenamefont {Omran}, \citenamefont {Pichler}, \citenamefont
  {Choi}, \citenamefont {Zibrov}, \citenamefont {Endres}, \citenamefont
  {Greiner}, \citenamefont {Vuleti{\'c}},\ and\ \citenamefont
  {Lukin}}]{Bernien2017Dynamics}%
  \BibitemOpen
  \bibfield  {author} {\bibinfo {author} {\bibfnamefont {H.}~\bibnamefont
  {Bernien}}, \bibinfo {author} {\bibfnamefont {S.}~\bibnamefont {Schwartz}},
  \bibinfo {author} {\bibfnamefont {A.}~\bibnamefont {Keesling}}, \bibinfo
  {author} {\bibfnamefont {H.}~\bibnamefont {Levine}}, \bibinfo {author}
  {\bibfnamefont {A.}~\bibnamefont {Omran}}, \bibinfo {author} {\bibfnamefont
  {H.}~\bibnamefont {Pichler}}, \bibinfo {author} {\bibfnamefont
  {S.}~\bibnamefont {Choi}}, \bibinfo {author} {\bibfnamefont {A.~S.}\
  \bibnamefont {Zibrov}}, \bibinfo {author} {\bibfnamefont {M.}~\bibnamefont
  {Endres}}, \bibinfo {author} {\bibfnamefont {M.}~\bibnamefont {Greiner}},
  \bibinfo {author} {\bibfnamefont {V.}~\bibnamefont {Vuleti{\'c}}},\ and\
  \bibinfo {author} {\bibfnamefont {M.~D.}\ \bibnamefont {Lukin}},\ }\bibfield
  {title} {\bibinfo {title} {Probing many-body dynamics on a 51-atom quantum
  simulator},\ }\href {https://doi.org/10.1038/nature24622} {\bibfield
  {journal} {\bibinfo  {journal} {Nature}\ }\textbf {\bibinfo {volume} {551}},\
  \bibinfo {pages} {579} (\bibinfo {year} {2017})}\BibitemShut {NoStop}%
\bibitem [{\citenamefont {Keesling}\ \emph {et~al.}(2019)\citenamefont
  {Keesling}, \citenamefont {Omran}, \citenamefont {Levine}, \citenamefont
  {Bernien}, \citenamefont {Pichler}, \citenamefont {Choi}, \citenamefont
  {Samajdar}, \citenamefont {Schwartz}, \citenamefont {Silvi}, \citenamefont
  {Sachdev}, \citenamefont {Zoller}, \citenamefont {Endres}, \citenamefont
  {Greiner}, \citenamefont {Vuleti{\'c}},\ and\ \citenamefont
  {Lukin}}]{Keesling2019Kibble}%
  \BibitemOpen
  \bibfield  {author} {\bibinfo {author} {\bibfnamefont {A.}~\bibnamefont
  {Keesling}}, \bibinfo {author} {\bibfnamefont {A.}~\bibnamefont {Omran}},
  \bibinfo {author} {\bibfnamefont {H.}~\bibnamefont {Levine}}, \bibinfo
  {author} {\bibfnamefont {H.}~\bibnamefont {Bernien}}, \bibinfo {author}
  {\bibfnamefont {H.}~\bibnamefont {Pichler}}, \bibinfo {author} {\bibfnamefont
  {S.}~\bibnamefont {Choi}}, \bibinfo {author} {\bibfnamefont {R.}~\bibnamefont
  {Samajdar}}, \bibinfo {author} {\bibfnamefont {S.}~\bibnamefont {Schwartz}},
  \bibinfo {author} {\bibfnamefont {P.}~\bibnamefont {Silvi}}, \bibinfo
  {author} {\bibfnamefont {S.}~\bibnamefont {Sachdev}}, \bibinfo {author}
  {\bibfnamefont {P.}~\bibnamefont {Zoller}}, \bibinfo {author} {\bibfnamefont
  {M.}~\bibnamefont {Endres}}, \bibinfo {author} {\bibfnamefont
  {M.}~\bibnamefont {Greiner}}, \bibinfo {author} {\bibfnamefont
  {V.}~\bibnamefont {Vuleti{\'c}}},\ and\ \bibinfo {author} {\bibfnamefont
  {M.~D.}\ \bibnamefont {Lukin}},\ }\bibfield  {title} {\bibinfo {title}
  {Quantum kibble--zurek mechanism and critical dynamics on a programmable
  rydberg simulator},\ }\href {https://doi.org/10.1038/s41586-019-1070-1}
  {\bibfield  {journal} {\bibinfo  {journal} {Nature}\ }\textbf {\bibinfo
  {volume} {568}},\ \bibinfo {pages} {207} (\bibinfo {year}
  {2019})}\BibitemShut {NoStop}%
\bibitem [{\citenamefont {Labuhn}\ \emph {et~al.}(2016)\citenamefont {Labuhn},
  \citenamefont {Barredo}, \citenamefont {Ravets}, \citenamefont
  {de~L{\'e}s{\'e}leuc}, \citenamefont {Macr{\`\i}}, \citenamefont {Lahaye},\
  and\ \citenamefont {Browaeys}}]{Labuhn2016RydIsing}%
  \BibitemOpen
  \bibfield  {author} {\bibinfo {author} {\bibfnamefont {H.}~\bibnamefont
  {Labuhn}}, \bibinfo {author} {\bibfnamefont {D.}~\bibnamefont {Barredo}},
  \bibinfo {author} {\bibfnamefont {S.}~\bibnamefont {Ravets}}, \bibinfo
  {author} {\bibfnamefont {S.}~\bibnamefont {de~L{\'e}s{\'e}leuc}}, \bibinfo
  {author} {\bibfnamefont {T.}~\bibnamefont {Macr{\`\i}}}, \bibinfo {author}
  {\bibfnamefont {T.}~\bibnamefont {Lahaye}},\ and\ \bibinfo {author}
  {\bibfnamefont {A.}~\bibnamefont {Browaeys}},\ }\bibfield  {title} {\bibinfo
  {title} {Tunable two-dimensional arrays of single rydberg atoms for realizing
  quantum ising models},\ }\href {https://doi.org/10.1038/nature18274}
  {\bibfield  {journal} {\bibinfo  {journal} {Nature}\ }\textbf {\bibinfo
  {volume} {534}},\ \bibinfo {pages} {667} (\bibinfo {year}
  {2016})}\BibitemShut {NoStop}%
\bibitem [{\citenamefont {de~Léséleuc}\ \emph {et~al.}(2019)\citenamefont
  {de~Léséleuc}, \citenamefont {Lienhard}, \citenamefont {Scholl},
  \citenamefont {Barredo}, \citenamefont {Weber}, \citenamefont {Lang},
  \citenamefont {Büchler}, \citenamefont {Lahaye},\ and\ \citenamefont
  {Browaeys}}]{Leseleuc2019topo}%
  \BibitemOpen
  \bibfield  {author} {\bibinfo {author} {\bibfnamefont {S.}~\bibnamefont
  {de~Léséleuc}}, \bibinfo {author} {\bibfnamefont {V.}~\bibnamefont
  {Lienhard}}, \bibinfo {author} {\bibfnamefont {P.}~\bibnamefont {Scholl}},
  \bibinfo {author} {\bibfnamefont {D.}~\bibnamefont {Barredo}}, \bibinfo
  {author} {\bibfnamefont {S.}~\bibnamefont {Weber}}, \bibinfo {author}
  {\bibfnamefont {N.}~\bibnamefont {Lang}}, \bibinfo {author} {\bibfnamefont
  {H.~P.}\ \bibnamefont {Büchler}}, \bibinfo {author} {\bibfnamefont
  {T.}~\bibnamefont {Lahaye}},\ and\ \bibinfo {author} {\bibfnamefont
  {A.}~\bibnamefont {Browaeys}},\ }\bibfield  {title} {\bibinfo {title}
  {Observation of a symmetry-protected topological phase of interacting bosons
  with rydberg atoms},\ }\href {https://doi.org/10.1126/science.aav9105}
  {\bibfield  {journal} {\bibinfo  {journal} {Science}\ }\textbf {\bibinfo
  {volume} {365}},\ \bibinfo {pages} {775} (\bibinfo {year}
  {2019})}\BibitemShut {NoStop}%
\bibitem [{\citenamefont {Ebadi}\ \emph {et~al.}(2021)\citenamefont {Ebadi},
  \citenamefont {Wang}, \citenamefont {Levine}, \citenamefont {Keesling},
  \citenamefont {Semeghini}, \citenamefont {Omran}, \citenamefont {Bluvstein},
  \citenamefont {Samajdar}, \citenamefont {Pichler}, \citenamefont {Ho},
  \citenamefont {Choi}, \citenamefont {Sachdev}, \citenamefont {Greiner},
  \citenamefont {Vuleti{\'c}},\ and\ \citenamefont {Lukin}}]{Ebadi2021_256}%
  \BibitemOpen
  \bibfield  {author} {\bibinfo {author} {\bibfnamefont {S.}~\bibnamefont
  {Ebadi}}, \bibinfo {author} {\bibfnamefont {T.~T.}\ \bibnamefont {Wang}},
  \bibinfo {author} {\bibfnamefont {H.}~\bibnamefont {Levine}}, \bibinfo
  {author} {\bibfnamefont {A.}~\bibnamefont {Keesling}}, \bibinfo {author}
  {\bibfnamefont {G.}~\bibnamefont {Semeghini}}, \bibinfo {author}
  {\bibfnamefont {A.}~\bibnamefont {Omran}}, \bibinfo {author} {\bibfnamefont
  {D.}~\bibnamefont {Bluvstein}}, \bibinfo {author} {\bibfnamefont
  {R.}~\bibnamefont {Samajdar}}, \bibinfo {author} {\bibfnamefont
  {H.}~\bibnamefont {Pichler}}, \bibinfo {author} {\bibfnamefont {W.~W.}\
  \bibnamefont {Ho}}, \bibinfo {author} {\bibfnamefont {S.}~\bibnamefont
  {Choi}}, \bibinfo {author} {\bibfnamefont {S.}~\bibnamefont {Sachdev}},
  \bibinfo {author} {\bibfnamefont {M.}~\bibnamefont {Greiner}}, \bibinfo
  {author} {\bibfnamefont {V.}~\bibnamefont {Vuleti{\'c}}},\ and\ \bibinfo
  {author} {\bibfnamefont {M.~D.}\ \bibnamefont {Lukin}},\ }\bibfield  {title}
  {\bibinfo {title} {Quantum phases of matter on a 256-atom programmable
  quantum simulator},\ }\href {https://doi.org/10.1038/s41586-021-03582-4}
  {\bibfield  {journal} {\bibinfo  {journal} {Nature}\ }\textbf {\bibinfo
  {volume} {595}},\ \bibinfo {pages} {227} (\bibinfo {year}
  {2021})}\BibitemShut {NoStop}%
\bibitem [{\citenamefont {Scholl}\ \emph {et~al.}(2021)\citenamefont {Scholl},
  \citenamefont {Schuler}, \citenamefont {Williams}, \citenamefont
  {Eberharter}, \citenamefont {Barredo}, \citenamefont {Schymik}, \citenamefont
  {Lienhard}, \citenamefont {Henry}, \citenamefont {Lang}, \citenamefont
  {Lahaye}, \citenamefont {L{\"a}uchli},\ and\ \citenamefont
  {Browaeys}}]{Pascal2021AF}%
  \BibitemOpen
  \bibfield  {author} {\bibinfo {author} {\bibfnamefont {P.}~\bibnamefont
  {Scholl}}, \bibinfo {author} {\bibfnamefont {M.}~\bibnamefont {Schuler}},
  \bibinfo {author} {\bibfnamefont {H.~J.}\ \bibnamefont {Williams}}, \bibinfo
  {author} {\bibfnamefont {A.~A.}\ \bibnamefont {Eberharter}}, \bibinfo
  {author} {\bibfnamefont {D.}~\bibnamefont {Barredo}}, \bibinfo {author}
  {\bibfnamefont {K.-N.}\ \bibnamefont {Schymik}}, \bibinfo {author}
  {\bibfnamefont {V.}~\bibnamefont {Lienhard}}, \bibinfo {author}
  {\bibfnamefont {L.-P.}\ \bibnamefont {Henry}}, \bibinfo {author}
  {\bibfnamefont {T.~C.}\ \bibnamefont {Lang}}, \bibinfo {author}
  {\bibfnamefont {T.}~\bibnamefont {Lahaye}}, \bibinfo {author} {\bibfnamefont
  {A.~M.}\ \bibnamefont {L{\"a}uchli}},\ and\ \bibinfo {author} {\bibfnamefont
  {A.}~\bibnamefont {Browaeys}},\ }\bibfield  {title} {\bibinfo {title}
  {Quantum simulation of 2d antiferromagnets with hundreds of rydberg atoms},\
  }\href {https://doi.org/10.1038/s41586-021-03585-1} {\bibfield  {journal}
  {\bibinfo  {journal} {Nature}\ }\textbf {\bibinfo {volume} {595}},\ \bibinfo
  {pages} {233} (\bibinfo {year} {2021})}\BibitemShut {NoStop}%
\bibitem [{\citenamefont {Zhang}\ \emph {et~al.}(2025)\citenamefont {Zhang}
  \emph {et~al.}}]{floating}%
  \BibitemOpen
  \bibfield  {author} {\bibinfo {author} {\bibfnamefont {J.}~\bibnamefont
  {Zhang}} \emph {et~al.},\ }\bibfield  {title} {\bibinfo {title} {{Probing
  quantum floating phases in Rydberg atom arrays}},\ }\href
  {https://doi.org/10.1038/s41467-025-55947-2} {\bibfield  {journal} {\bibinfo
  {journal} {Nature Commun.}\ }\textbf {\bibinfo {volume} {16}},\ \bibinfo
  {pages} {712} (\bibinfo {year} {2025})},\ \Eprint
  {https://arxiv.org/abs/2401.08087} {arXiv:2401.08087 [quant-ph]} \BibitemShut
  {NoStop}%
\bibitem [{\citenamefont {Meurice}(2024)}]{SM}%
  \BibitemOpen
  \bibfield  {author} {\bibinfo {author} {\bibfnamefont {Y.}~\bibnamefont
  {Meurice}},\ }\href@noop {} {\bibinfo {title} {Supplementary material}}
  (\bibinfo {year} {2024})\BibitemShut {NoStop}%
\bibitem [{zen(2025)}]{zenodo}%
  \BibitemOpen
  \href@noop {} {\bibinfo {title} {Zenodo repository}},\ \bibinfo
  {howpublished} {\url{https://zenodo.org/records/15103785}} (\bibinfo {year}
  {2025})\BibitemShut {NoStop}%
\bibitem [{Note1()}]{Note1}%
  \BibitemOpen
  \bibinfo {note} {This was pointed out to us by Alex Lukin.}\BibitemShut
  {Stop}%
\bibitem [{\citenamefont {Lukin}\ \emph {et~al.}(2019)\citenamefont {Lukin},
  \citenamefont {Rispoli}, \citenamefont {Schittko}, \citenamefont {Tai},
  \citenamefont {Kaufman}, \citenamefont {Choi}, \citenamefont {Khemani},
  \citenamefont {Léonard},\ and\ \citenamefont {Greiner}}]{Lukin_2019}%
  \BibitemOpen
  \bibfield  {author} {\bibinfo {author} {\bibfnamefont {A.}~\bibnamefont
  {Lukin}}, \bibinfo {author} {\bibfnamefont {M.}~\bibnamefont {Rispoli}},
  \bibinfo {author} {\bibfnamefont {R.}~\bibnamefont {Schittko}}, \bibinfo
  {author} {\bibfnamefont {M.~E.}\ \bibnamefont {Tai}}, \bibinfo {author}
  {\bibfnamefont {A.~M.}\ \bibnamefont {Kaufman}}, \bibinfo {author}
  {\bibfnamefont {S.}~\bibnamefont {Choi}}, \bibinfo {author} {\bibfnamefont
  {V.}~\bibnamefont {Khemani}}, \bibinfo {author} {\bibfnamefont
  {J.}~\bibnamefont {Léonard}},\ and\ \bibinfo {author} {\bibfnamefont
  {M.}~\bibnamefont {Greiner}},\ }\bibfield  {title} {\bibinfo {title} {Probing
  entanglement in a many-body–localized system},\ }\href
  {https://doi.org/10.1126/science.aau0818} {\bibfield  {journal} {\bibinfo
  {journal} {Science}\ }\textbf {\bibinfo {volume} {364}},\ \bibinfo {pages}
  {256–260} (\bibinfo {year} {2019})}\BibitemShut {NoStop}%
\bibitem [{\citenamefont {Torlai}\ \emph {et~al.}(2019)\citenamefont {Torlai},
  \citenamefont {Timar}, \citenamefont {van Nieuwenburg}, \citenamefont
  {Levine}, \citenamefont {Omran}, \citenamefont {Keesling}, \citenamefont
  {Bernien}, \citenamefont {Greiner}, \citenamefont {Vuletić}, \citenamefont
  {Lukin}, \citenamefont {Melko},\ and\ \citenamefont {Endres}}]{Torlai_2019}%
  \BibitemOpen
  \bibfield  {author} {\bibinfo {author} {\bibfnamefont {G.}~\bibnamefont
  {Torlai}}, \bibinfo {author} {\bibfnamefont {B.}~\bibnamefont {Timar}},
  \bibinfo {author} {\bibfnamefont {E.~P.}\ \bibnamefont {van Nieuwenburg}},
  \bibinfo {author} {\bibfnamefont {H.}~\bibnamefont {Levine}}, \bibinfo
  {author} {\bibfnamefont {A.}~\bibnamefont {Omran}}, \bibinfo {author}
  {\bibfnamefont {A.}~\bibnamefont {Keesling}}, \bibinfo {author}
  {\bibfnamefont {H.}~\bibnamefont {Bernien}}, \bibinfo {author} {\bibfnamefont
  {M.}~\bibnamefont {Greiner}}, \bibinfo {author} {\bibfnamefont
  {V.}~\bibnamefont {Vuletić}}, \bibinfo {author} {\bibfnamefont {M.~D.}\
  \bibnamefont {Lukin}}, \bibinfo {author} {\bibfnamefont {R.~G.}\ \bibnamefont
  {Melko}},\ and\ \bibinfo {author} {\bibfnamefont {M.}~\bibnamefont
  {Endres}},\ }\bibfield  {title} {\bibinfo {title} {Integrating neural
  networks with a quantum simulator for state reconstruction},\ }\bibfield
  {journal} {\bibinfo  {journal} {Physical Review Letters}\ }\textbf {\bibinfo
  {volume} {123}},\ \href {https://doi.org/10.1103/physrevlett.123.230504}
  {10.1103/physrevlett.123.230504} (\bibinfo {year} {2019})\BibitemShut
  {NoStop}%
\bibitem [{\citenamefont {Zhang}\ \emph {et~al.}(2018)\citenamefont {Zhang},
  \citenamefont {Unmuth-Yockey}, \citenamefont {Zeiher}, \citenamefont
  {Bazavov}, \citenamefont {Tsai},\ and\ \citenamefont
  {Meurice}}]{Zhang:2018ufj}%
  \BibitemOpen
  \bibfield  {author} {\bibinfo {author} {\bibfnamefont {J.}~\bibnamefont
  {Zhang}}, \bibinfo {author} {\bibfnamefont {J.}~\bibnamefont
  {Unmuth-Yockey}}, \bibinfo {author} {\bibfnamefont {J.}~\bibnamefont
  {Zeiher}}, \bibinfo {author} {\bibfnamefont {A.}~\bibnamefont {Bazavov}},
  \bibinfo {author} {\bibfnamefont {S.-W.}\ \bibnamefont {Tsai}},\ and\
  \bibinfo {author} {\bibfnamefont {Y.}~\bibnamefont {Meurice}},\ }\bibfield
  {title} {\bibinfo {title} {Quantum simulation of the universal features of
  the polyakov loop},\ }\href {https://doi.org/10.1103/PhysRevLett.121.223201}
  {\bibfield  {journal} {\bibinfo  {journal} {Phys. Rev. Lett.}\ }\textbf
  {\bibinfo {volume} {121}},\ \bibinfo {pages} {223201} (\bibinfo {year}
  {2018})}\BibitemShut {NoStop}%
\bibitem [{\citenamefont {Surace}\ \emph {et~al.}(2020)\citenamefont {Surace},
  \citenamefont {Mazza}, \citenamefont {Giudici}, \citenamefont {Lerose},
  \citenamefont {Gambassi},\ and\ \citenamefont {Dalmonte}}]{Surace:2019dtp}%
  \BibitemOpen
  \bibfield  {author} {\bibinfo {author} {\bibfnamefont {F.~M.}\ \bibnamefont
  {Surace}}, \bibinfo {author} {\bibfnamefont {P.~P.}\ \bibnamefont {Mazza}},
  \bibinfo {author} {\bibfnamefont {G.}~\bibnamefont {Giudici}}, \bibinfo
  {author} {\bibfnamefont {A.}~\bibnamefont {Lerose}}, \bibinfo {author}
  {\bibfnamefont {A.}~\bibnamefont {Gambassi}},\ and\ \bibinfo {author}
  {\bibfnamefont {M.}~\bibnamefont {Dalmonte}},\ }\bibfield  {title} {\bibinfo
  {title} {Lattice gauge theories and string dynamics in rydberg atom quantum
  simulators},\ }\href {https://doi.org/10.1103/PhysRevX.10.021041} {\bibfield
  {journal} {\bibinfo  {journal} {Phys. Rev. X}\ }\textbf {\bibinfo {volume}
  {10}},\ \bibinfo {pages} {021041} (\bibinfo {year} {2020})}\BibitemShut
  {NoStop}%
\bibitem [{\citenamefont {Notarnicola}\ \emph {et~al.}(2020)\citenamefont
  {Notarnicola}, \citenamefont {Collura},\ and\ \citenamefont
  {Montangero}}]{Notarnicola:2019wzb}%
  \BibitemOpen
  \bibfield  {author} {\bibinfo {author} {\bibfnamefont {S.}~\bibnamefont
  {Notarnicola}}, \bibinfo {author} {\bibfnamefont {M.}~\bibnamefont
  {Collura}},\ and\ \bibinfo {author} {\bibfnamefont {S.}~\bibnamefont
  {Montangero}},\ }\bibfield  {title} {\bibinfo {title} {Real-time-dynamics
  quantum simulation of $(1+1)\text{-dimensional}$ lattice qed with rydberg
  atoms},\ }\href {https://doi.org/10.1103/PhysRevResearch.2.013288} {\bibfield
   {journal} {\bibinfo  {journal} {Phys. Rev. Res.}\ }\textbf {\bibinfo
  {volume} {2}},\ \bibinfo {pages} {013288} (\bibinfo {year}
  {2020})}\BibitemShut {NoStop}%
\bibitem [{\citenamefont {Celi}\ \emph {et~al.}(2020)\citenamefont {Celi},
  \citenamefont {Vermersch}, \citenamefont {Viyuela}, \citenamefont {Pichler},
  \citenamefont {Lukin},\ and\ \citenamefont {Zoller}}]{Celi:2019lqy}%
  \BibitemOpen
  \bibfield  {author} {\bibinfo {author} {\bibfnamefont {A.}~\bibnamefont
  {Celi}}, \bibinfo {author} {\bibfnamefont {B.}~\bibnamefont {Vermersch}},
  \bibinfo {author} {\bibfnamefont {O.}~\bibnamefont {Viyuela}}, \bibinfo
  {author} {\bibfnamefont {H.}~\bibnamefont {Pichler}}, \bibinfo {author}
  {\bibfnamefont {M.~D.}\ \bibnamefont {Lukin}},\ and\ \bibinfo {author}
  {\bibfnamefont {P.}~\bibnamefont {Zoller}},\ }\bibfield  {title} {\bibinfo
  {title} {{Emerging Two-Dimensional Gauge Theories in Rydberg Configurable
  Arrays}},\ }\href {https://doi.org/10.1103/PhysRevX.10.021057} {\bibfield
  {journal} {\bibinfo  {journal} {Phys. Rev. X}\ }\textbf {\bibinfo {volume}
  {10}},\ \bibinfo {pages} {021057} (\bibinfo {year} {2020})},\ \Eprint
  {https://arxiv.org/abs/1907.03311} {arXiv:1907.03311 [quant-ph]} \BibitemShut
  {NoStop}%
\bibitem [{\citenamefont {Meurice}(2021)}]{Meurice:2021pvj}%
  \BibitemOpen
  \bibfield  {author} {\bibinfo {author} {\bibfnamefont {Y.}~\bibnamefont
  {Meurice}},\ }\bibfield  {title} {\bibinfo {title} {{Theoretical methods to
  design and test quantum simulators for the compact Abelian Higgs model}},\
  }\href {https://doi.org/10.1103/PhysRevD.104.094513} {\bibfield  {journal}
  {\bibinfo  {journal} {Phys. Rev. D}\ }\textbf {\bibinfo {volume} {104}},\
  \bibinfo {pages} {094513} (\bibinfo {year} {2021})},\ \Eprint
  {https://arxiv.org/abs/2107.11366} {arXiv:2107.11366 [quant-ph]} \BibitemShut
  {NoStop}%
\bibitem [{\citenamefont {Fromholz}\ \emph {et~al.}(2022)\citenamefont
  {Fromholz}, \citenamefont {Tsitsishvili}, \citenamefont {Votto},
  \citenamefont {Dalmonte}, \citenamefont {Nersesyan},\ and\ \citenamefont
  {Chanda}}]{Fromholz:2022ymy}%
  \BibitemOpen
  \bibfield  {author} {\bibinfo {author} {\bibfnamefont {P.}~\bibnamefont
  {Fromholz}}, \bibinfo {author} {\bibfnamefont {M.}~\bibnamefont
  {Tsitsishvili}}, \bibinfo {author} {\bibfnamefont {M.}~\bibnamefont {Votto}},
  \bibinfo {author} {\bibfnamefont {M.}~\bibnamefont {Dalmonte}}, \bibinfo
  {author} {\bibfnamefont {A.}~\bibnamefont {Nersesyan}},\ and\ \bibinfo
  {author} {\bibfnamefont {T.}~\bibnamefont {Chanda}},\ }\bibfield  {title}
  {\bibinfo {title} {Phase diagram of rydberg-dressed atoms on two-leg
  triangular ladders},\ }\href {https://doi.org/10.1103/PhysRevB.106.155411}
  {\bibfield  {journal} {\bibinfo  {journal} {Phys. Rev. B}\ }\textbf {\bibinfo
  {volume} {106}},\ \bibinfo {pages} {155411} (\bibinfo {year}
  {2022})}\BibitemShut {NoStop}%
\bibitem [{\citenamefont {Gonz\'alez-Cuadra}\ \emph {et~al.}(2022)\citenamefont
  {Gonz\'alez-Cuadra}, \citenamefont {Zache}, \citenamefont {Carrasco},
  \citenamefont {Kraus},\ and\ \citenamefont
  {Zoller}}]{Gonzalez-Cuadra:2022hxt}%
  \BibitemOpen
  \bibfield  {author} {\bibinfo {author} {\bibfnamefont {D.}~\bibnamefont
  {Gonz\'alez-Cuadra}}, \bibinfo {author} {\bibfnamefont {T.~V.}\ \bibnamefont
  {Zache}}, \bibinfo {author} {\bibfnamefont {J.}~\bibnamefont {Carrasco}},
  \bibinfo {author} {\bibfnamefont {B.}~\bibnamefont {Kraus}},\ and\ \bibinfo
  {author} {\bibfnamefont {P.}~\bibnamefont {Zoller}},\ }\bibfield  {title}
  {\bibinfo {title} {{Hardware Efficient Quantum Simulation of Non-Abelian
  Gauge Theories with Qudits on Rydberg Platforms}},\ }\href
  {https://doi.org/10.1103/PhysRevLett.129.160501} {\bibfield  {journal}
  {\bibinfo  {journal} {Phys. Rev. Lett.}\ }\textbf {\bibinfo {volume} {129}},\
  \bibinfo {pages} {160501} (\bibinfo {year} {2022})},\ \Eprint
  {https://arxiv.org/abs/2203.15541} {arXiv:2203.15541 [quant-ph]} \BibitemShut
  {NoStop}%
\bibitem [{\citenamefont {Heitritter}\ \emph {et~al.}(2022)\citenamefont
  {Heitritter}, \citenamefont {Meurice},\ and\ \citenamefont
  {Mrenna}}]{Heitritter:2022jik}%
  \BibitemOpen
  \bibfield  {author} {\bibinfo {author} {\bibfnamefont {K.}~\bibnamefont
  {Heitritter}}, \bibinfo {author} {\bibfnamefont {Y.}~\bibnamefont
  {Meurice}},\ and\ \bibinfo {author} {\bibfnamefont {S.}~\bibnamefont
  {Mrenna}},\ }\href@noop {} {\bibinfo {title} {{Prolegomena to a hybrid
  Classical/Rydberg simulator for hadronization (QuPYTH)}}} (\bibinfo {year}
  {2022}),\ \Eprint {https://arxiv.org/abs/2212.02476} {arXiv:2212.02476
  [quant-ph]} \BibitemShut {NoStop}%
\bibitem [{\citenamefont {Bauer}\ \emph {et~al.}(2023)\citenamefont {Bauer}
  \emph {et~al.}}]{Bauer:2022hpo}%
  \BibitemOpen
  \bibfield  {author} {\bibinfo {author} {\bibfnamefont {C.~W.}\ \bibnamefont
  {Bauer}} \emph {et~al.},\ }\bibfield  {title} {\bibinfo {title} {{Quantum
  Simulation for High-Energy Physics}},\ }\href
  {https://doi.org/10.1103/PRXQuantum.4.027001} {\bibfield  {journal} {\bibinfo
   {journal} {PRX Quantum}\ }\textbf {\bibinfo {volume} {4}},\ \bibinfo {pages}
  {027001} (\bibinfo {year} {2023})},\ \Eprint
  {https://arxiv.org/abs/2204.03381} {arXiv:2204.03381 [quant-ph]} \BibitemShut
  {NoStop}%
\bibitem [{\citenamefont {Halimeh}\ \emph {et~al.}(2023)\citenamefont
  {Halimeh}, \citenamefont {Aidelsburger}, \citenamefont {Grusdt},
  \citenamefont {Hauke},\ and\ \citenamefont {Yang}}]{Halimeh:2023lid}%
  \BibitemOpen
  \bibfield  {author} {\bibinfo {author} {\bibfnamefont {J.~C.}\ \bibnamefont
  {Halimeh}}, \bibinfo {author} {\bibfnamefont {M.}~\bibnamefont
  {Aidelsburger}}, \bibinfo {author} {\bibfnamefont {F.}~\bibnamefont
  {Grusdt}}, \bibinfo {author} {\bibfnamefont {P.}~\bibnamefont {Hauke}},\ and\
  \bibinfo {author} {\bibfnamefont {B.}~\bibnamefont {Yang}},\ }\href@noop {}
  {\bibinfo {title} {{Cold-atom quantum simulators of gauge theories}}}
  (\bibinfo {year} {2023}),\ \Eprint {https://arxiv.org/abs/2310.12201}
  {arXiv:2310.12201 [cond-mat.quant-gas]} \BibitemShut {NoStop}%
\bibitem [{\citenamefont {Zhang}\ \emph {et~al.}(2024)\citenamefont {Zhang},
  \citenamefont {Tsai},\ and\ \citenamefont {Meurice}}]{Zhang:2023agx}%
  \BibitemOpen
  \bibfield  {author} {\bibinfo {author} {\bibfnamefont {J.}~\bibnamefont
  {Zhang}}, \bibinfo {author} {\bibfnamefont {S.~W.}\ \bibnamefont {Tsai}},\
  and\ \bibinfo {author} {\bibfnamefont {Y.}~\bibnamefont {Meurice}},\
  }\bibfield  {title} {\bibinfo {title} {{Critical behavior of lattice gauge
  theory Rydberg simulators from effective Hamiltonians}},\ }\href
  {https://doi.org/10.1103/PhysRevD.110.034513} {\bibfield  {journal} {\bibinfo
   {journal} {Phys. Rev. D}\ }\textbf {\bibinfo {volume} {110}},\ \bibinfo
  {pages} {034513} (\bibinfo {year} {2024})},\ \Eprint
  {https://arxiv.org/abs/2312.04436} {arXiv:2312.04436 [quant-ph]} \BibitemShut
  {NoStop}%
\bibitem [{\citenamefont {Pisarski}\ \emph {et~al.}(2019)\citenamefont
  {Pisarski}, \citenamefont {Skokov},\ and\ \citenamefont
  {Tsvelik}}]{Pisarski:2019cvo}%
  \BibitemOpen
  \bibfield  {author} {\bibinfo {author} {\bibfnamefont {R.~D.}\ \bibnamefont
  {Pisarski}}, \bibinfo {author} {\bibfnamefont {V.~V.}\ \bibnamefont
  {Skokov}},\ and\ \bibinfo {author} {\bibfnamefont {A.~M.}\ \bibnamefont
  {Tsvelik}},\ }\bibfield  {title} {\bibinfo {title} {{A Pedagogical
  Introduction to the Lifshitz Regime}},\ }\href
  {https://doi.org/10.3390/universe5020048} {\bibfield  {journal} {\bibinfo
  {journal} {Universe}\ }\textbf {\bibinfo {volume} {5}},\ \bibinfo {pages}
  {48} (\bibinfo {year} {2019})}\BibitemShut {NoStop}%
\bibitem [{\citenamefont {Kojo}\ \emph {et~al.}(2010)\citenamefont {Kojo},
  \citenamefont {Hidaka}, \citenamefont {McLerran},\ and\ \citenamefont
  {Pisarski}}]{Kojo:2009ha}%
  \BibitemOpen
  \bibfield  {author} {\bibinfo {author} {\bibfnamefont {T.}~\bibnamefont
  {Kojo}}, \bibinfo {author} {\bibfnamefont {Y.}~\bibnamefont {Hidaka}},
  \bibinfo {author} {\bibfnamefont {L.}~\bibnamefont {McLerran}},\ and\
  \bibinfo {author} {\bibfnamefont {R.~D.}\ \bibnamefont {Pisarski}},\
  }\bibfield  {title} {\bibinfo {title} {Quarkyonic chiral spirals},\ }\href
  {https://doi.org/https://doi.org/10.1016/j.nuclphysa.2010.05.053} {\bibfield
  {journal} {\bibinfo  {journal} {Nuclear Physics A}\ }\textbf {\bibinfo
  {volume} {843}},\ \bibinfo {pages} {37} (\bibinfo {year} {2010})}\BibitemShut
  {NoStop}%
\bibitem [{\citenamefont {Headrick}\ and\ \citenamefont
  {Takayanagi}(2007)}]{Headrick:2007km}%
  \BibitemOpen
  \bibfield  {author} {\bibinfo {author} {\bibfnamefont {M.}~\bibnamefont
  {Headrick}}\ and\ \bibinfo {author} {\bibfnamefont {T.}~\bibnamefont
  {Takayanagi}},\ }\bibfield  {title} {\bibinfo {title} {{A Holographic proof
  of the strong subadditivity of entanglement entropy}},\ }\href
  {https://doi.org/10.1103/PhysRevD.76.106013} {\bibfield  {journal} {\bibinfo
  {journal} {Phys. Rev. D}\ }\textbf {\bibinfo {volume} {76}},\ \bibinfo
  {pages} {106013} (\bibinfo {year} {2007})},\ \Eprint
  {https://arxiv.org/abs/0704.3719} {arXiv:0704.3719 [hep-th]} \BibitemShut
  {NoStop}%
\bibitem [{\citenamefont {Hayden}\ \emph {et~al.}(2013)\citenamefont {Hayden},
  \citenamefont {Headrick},\ and\ \citenamefont {Maloney}}]{Hayden:2011ag}%
  \BibitemOpen
  \bibfield  {author} {\bibinfo {author} {\bibfnamefont {P.}~\bibnamefont
  {Hayden}}, \bibinfo {author} {\bibfnamefont {M.}~\bibnamefont {Headrick}},\
  and\ \bibinfo {author} {\bibfnamefont {A.}~\bibnamefont {Maloney}},\
  }\bibfield  {title} {\bibinfo {title} {{Holographic Mutual Information is
  Monogamous}},\ }\href {https://doi.org/10.1103/PhysRevD.87.046003} {\bibfield
   {journal} {\bibinfo  {journal} {Phys. Rev. D}\ }\textbf {\bibinfo {volume}
  {87}},\ \bibinfo {pages} {046003} (\bibinfo {year} {2013})},\ \Eprint
  {https://arxiv.org/abs/1107.2940} {arXiv:1107.2940 [hep-th]} \BibitemShut
  {NoStop}%
\bibitem [{\citenamefont {Lin}\ and\ \citenamefont
  {McGreevy}(2023)}]{Lin:2023pvl}%
  \BibitemOpen
  \bibfield  {author} {\bibinfo {author} {\bibfnamefont {T.-C.}\ \bibnamefont
  {Lin}}\ and\ \bibinfo {author} {\bibfnamefont {J.}~\bibnamefont {McGreevy}},\
  }\bibfield  {title} {\bibinfo {title} {{Conformal Field Theory Ground States
  as Critical Points of an Entropy Function}},\ }\href
  {https://doi.org/10.1103/PhysRevLett.131.251602} {\bibfield  {journal}
  {\bibinfo  {journal} {Phys. Rev. Lett.}\ }\textbf {\bibinfo {volume} {131}},\
  \bibinfo {pages} {251602} (\bibinfo {year} {2023})},\ \Eprint
  {https://arxiv.org/abs/2303.05444} {arXiv:2303.05444 [hep-th]} \BibitemShut
  {NoStop}%
\bibitem [{\citenamefont {Asaduzzaman}\ \emph {et~al.}(2024)\citenamefont
  {Asaduzzaman}, \citenamefont {Catterall}, \citenamefont {Maloney},
  \citenamefont {Meurice}, \citenamefont {Samlodia},\ and\ \citenamefont
  {Toga}}]{syrprogress}%
  \BibitemOpen
  \bibfield  {author} {\bibinfo {author} {\bibfnamefont {M.}~\bibnamefont
  {Asaduzzaman}}, \bibinfo {author} {\bibfnamefont {S.}~\bibnamefont
  {Catterall}}, \bibinfo {author} {\bibfnamefont {A.}~\bibnamefont {Maloney}},
  \bibinfo {author} {\bibfnamefont {Y.}~\bibnamefont {Meurice}}, \bibinfo
  {author} {\bibfnamefont {A.}~\bibnamefont {Samlodia}},\ and\ \bibinfo
  {author} {\bibfnamefont {G.~C.}\ \bibnamefont {Toga}},\ }\href@noop {}
  {\bibinfo {title} {Work in progress}} (\bibinfo {year} {2024})\BibitemShut
  {NoStop}%
\bibitem [{\citenamefont {Corona}\ \emph {et~al.}(2024)\citenamefont {Corona},
  \citenamefont {Asaduzzaman},\ and\ \citenamefont {Meurice}}]{jamesprogress}%
  \BibitemOpen
  \bibfield  {author} {\bibinfo {author} {\bibfnamefont {J.}~\bibnamefont
  {Corona}}, \bibinfo {author} {\bibfnamefont {M.}~\bibnamefont
  {Asaduzzaman}},\ and\ \bibinfo {author} {\bibfnamefont {Y.}~\bibnamefont
  {Meurice}},\ }\href@noop {} {\bibinfo {title} {Work in progress}} (\bibinfo
  {year} {2024})\BibitemShut {NoStop}%
\end{thebibliography}
%\end{document}

\end{document}